\documentclass[reprint,twocolumn,aps,pra,amsmath,amssymb,floatfix]{revtex4-1}
\usepackage[latin9]{inputenc}
\usepackage{amsmath}
\usepackage{amssymb}
\usepackage{graphicx}

\makeatletter
\usepackage{amssymb}
\usepackage{epsfig}
\usepackage{bm}
\usepackage{dcolumn}
\usepackage{color}
\usepackage{hyperref}
\hypersetup{
    colorlinks=true,
    linkcolor=blue,
    filecolor=magenta,
    urlcolor=cyan,
}
\makeatother

\begin{document}

\title{Quantum depletion and superfluid density of a supersolid in Raman spin-orbit-coupled Bose gases}

\author{Xiao-Long Chen}
\email{xiaolongchen@swin.edu.au}

\affiliation{Centre for Quantum and Optical Science, Swinburne University of Technology,
Melbourne, Victoria 3122, Australia}

\author{Jia Wang}

\affiliation{Centre for Quantum and Optical Science, Swinburne University of Technology,
Melbourne, Victoria 3122, Australia}

\author{Yun Li}

\affiliation{Centre for Quantum and Optical Science, Swinburne University of Technology,
Melbourne, Victoria 3122, Australia}

\author{Xia-Ji Liu}

\affiliation{Centre for Quantum and Optical Science, Swinburne University of Technology,
Melbourne, Victoria 3122, Australia}

\author{Hui Hu}

\affiliation{Centre for Quantum and Optical Science, Swinburne University of Technology,
Melbourne, Victoria 3122, Australia}

\date{\today}
\begin{abstract}
We theoretically investigate a three-dimensional weakly interacting Bose gas with one-dimensional Raman-type spin-orbit coupling at zero temperature. By employing an improved ansatz, including high-order harmonics in the stripe phase, we show that the critical transition from the stripe to the plane-wave phases is shifted to a relatively larger Rabi frequency compared to the prediction by previous work {[}Li \textit{et al.}, Phys. Rev. Lett. \textbf{108}, 225301 (2012){]} using a first-order stripe ansatz. We also determine the quantum depletion and superfluid density over a large range of Rabi frequency in different phases. The depletion exhibits an intriguing behavior with a discontinuous jump at the transition between the stripe and plane-wave phases, and a maximum at the transition between the plane-wave and zero-momentum phases. The superfluid density is derived through a phase-twist method. In the plane-wave and zero-momentum phases, it is significantly suppressed along the spin-orbit-coupling direction and vanishes at the transition, consistent with a recent work {[}Zhang \textit{et al.}, Phys. Rev. A \textbf{94}, 033635 (2016){]}, while in the stripe phase, it smoothly decreases with increasing Rabi frequency. Our predictions would be useful for further theoretical and experimental studies of the exotic supersolid stripe phase.
\end{abstract}

\maketitle

\section{Introduction}

The spin-orbit coupling (SOC) arising from the interaction of a particle's spin with its motion plays a crucial role in various fields of physics~\cite{Galitski2013,zhai2015degenerate}, intriguing plenty of studies on fascinating phenomena, such as the quantum spin Hall effect~\cite{zhu2006spin,liu2007optically,Beeler2013}, topological superfluidity~\cite{sato2009non,jiang2011majorana,liu2012probing,liu2012topological,*liu2013topological,liu2014realization} and exotic bosonic phases of matter~\cite{wang2010spin,ho2011bose,li2012quantum,hu2012spin}. In the last decade, this effect can be synthesized in ultracold neutral atoms~\cite{bloch2008many} through the atom-light interaction~\cite{liu2009effect,spielman2009raman}. The realization of Raman-laser-induced one-dimensional SOC~\cite{lin2011spin,wang2012spin,cheuk2012spin,dalibard2011colloquium,goldman2014light}, also referred to as equal Rashba and Dresselhaus SOC (ERDSOC), and two-dimensional SOC for both Bose~\cite{wu2016realization,sun2017long} and Fermi gases~\cite{huang2016experimental}, has provided a versatile platform for understanding the interplay between SOC and quantum many-body physics~\cite{li2012sum,martone2012anisotropic,zheng2013properties,zhang2012collective,ji2014experimental,ji2015softening}.

In a three-dimensional weakly interacting Bose gas with ERDSOC, the system exhibits sequentially three exotic condensation phases at zero temperature, i.e., the stripe (ST), plane-wave (PW), and zero-momentum (ZM) phases as the Rabi frequency of the Raman laser beams gradually rises~\cite{lin2011spin,li2012sum,martone2012anisotropic,zheng2013properties}. Previous investigations mainly focus on the last two phases. The problems that have been addressed include the ground-state phase diagram~\cite{li2012quantum,zheng2013properties}, quantum and thermal fluctuations~\cite{ozawa2012stability,cui2013enhancement,liao2014spin,chen2017quantum}, collective excitations~\cite{khamehchi2014measurement,chen2017collective}, superfluidity and critical velocities~\cite{zhu2012exotic,zhou2012opposite,yu2017landau}. However, only a handful of works involved the stripe phase, which attracts great attention after being directly observed in recent experiments with ultracold atomic gases~\cite{leonard2017supersolid,li2017stripe}.

Theoretically, the existence of a stripe phase was first predicted in Refs.~\cite{wang2010spin,ho2011bose,li2012quantum} using a first-order stripe ansatz. Later, by employing an improved high-order ansatz and calculating the static structure factor, one of the present authors (Y. L.) and her collaborators characterized the spin and density responses of the stripe phase and found two gapless modes in the elementary excitation spectrum~\cite{li2013superstripes}. This calculation clearly indicates the importance of the inclusion of high-order harmonics in the trial ansatz, for the purpose of \emph{quantitatively} characterizing the stripe phase. Unfortunately, apart from the elementary phonon excitation spectrum, none of the other physical properties of the stripe phase has so far been investigated using the ansatz with high-order harmonics.

Experimentally, in a SOC Bose gas of $^{87}$Rb atoms, the phase space for the stripe phase is small. Martone \textit{et al.} tried to enhance the stripe contrast and make it visible and stable under realistic experimental conditions, by theoretically considering the load of atoms into a two-dimensional bilayer configuration~\cite{martone2014approach}. Most recently, Li and co-workers achieved the effective SOC in optical superlattices and observed for the first time the exotic stripe phase with supersolid properties using Bragg spectroscopy~\cite{li2016spin,li2017stripe}. This experimental breakthrough provides a great opportunity to test and verify the theoretical predictions on the stripe phase.

In this work, motivated by previous theoretical studies and recent experiments, we explore the fascinating stripe phase and aim to make quantitative predictions of several fundamental properties of this phase using the high-order stripe ansatz. Within the Bogoliubov approximation, we first consider the dependence of the density distribution and the excitation spectrum on the tunable Rabi frequency. By introducing high-order harmonics and comparing the free energy in different trial ansatz, we then obtain an improved critical Rabi frequency for the transition from the stripe and plane-wave phases. We numerically calculate the depletion of the condensate induced by quantum fluctuations, with which one can straightforwardly characterize the first-order ST-PW transition and the second-order PW-ZM transition. Finally, by developing a phase-twist method, we discuss the superfluidity of a SOC Bose gas via calculating the superfluid density in all three phases.

This paper is organized as follows. We describe the model Hamiltonian and the theoretical framework in Sec.~\ref{sec:theory}. In Sec.~\ref{sec:results}, we present the density profile and the Bogoliubov excitation spectrum of the stripe phase (see Fig.~\ref{fig1}) and show how to improve the prediction on the critical ST-PW transition by taking high-order harmonics in the ansatz (Fig.~\ref{fig2} and~\ref{fig3}). We then calculate the quantum depletion as a function of Rabi frequency (Fig.~\ref{fig4}). The analytic expression of the superfluid density is derived for all the three phases using the first-order ansatz. In the stripe phase, the analytic prediction is compared with the more accurate numerical result with the high-order ansatz (see Fig.~\ref{fig5}). A summary and outlook are given in Sec.~\ref{sec:summary}.

\section{Theoretical Frameworks\label{sec:theory}}

\subsection{The model Hamiltonian}

We consider a three-dimensional weakly interacting Bose gas with Raman-induced spin-orbit coupling, the same as in our previous work~\cite{chen2017quantum}. The system can be described by a two-component (i.e., two-energy-level) Hamiltonian, $\hat{H}=\hat{H}_{0}+\hat{H}_{\mathrm{int}}$, where the single-particle Hamiltonian $\hat{H}_{0}$ and the interaction Hamiltonian $\hat{H}_{\mathrm{int}}$ read, respectively ($\hbar=1$)~\cite{lin2011spin,li2012quantum,zheng2013properties}
\begin{eqnarray}
\hat{H}_{0} & = & \int d^{3}{\bf r}[\hat{\Phi}_{\uparrow}^{\dagger}({\bf r}),\hat{\Phi}_{\downarrow}^{\dagger}({\bf r})]\mathcal{H}_{\mathrm{s}}(\hat{{\bf p}})\left[\begin{array}{c}
\hat{\Phi}_{\uparrow}({\bf r})\\
\hat{\Phi}_{\downarrow}({\bf r})
\end{array}\right],\label{eq:single-particle}\\
\hat{H}_{\mathrm{int}} & = & \int d^{3}{\bf r}\sum_{\sigma,\sigma^{\prime}=\uparrow,\downarrow}\frac{g_{\sigma\sigma^{\prime}}}{2}\hat{\Phi}_{\sigma}^{\dagger}({\bf r})\hat{\Phi}_{\sigma^{\prime}}^{\dagger}({\bf r})\hat{\Phi}_{\sigma^{\prime}}({\bf r})\hat{\Phi}_{\sigma}({\bf r}).\label{eq:interaction}
\end{eqnarray}
Here, $\mathcal{H}_{\mathrm{s}}(\hat{{\bf p}})$ is given by 
\begin{equation}
\mathcal{H}_{\mathrm{s}}(\hat{{\bf p}})=\frac{(\hat{{\bf p}}-k_{\mathrm{r}}\hat{{\bf e}}_{x}\sigma_{z})^{2}}{2m}+\frac{\Omega}{2}\sigma_{x}+\frac{\delta}{2}\sigma_{z},
\end{equation}
with the canonical momentum operator $\hat{{\bf p}}=-i\nabla$. It is worth noting that, in the Raman SOC scheme, the physical momenta of the two pseudo-spin states are $\hat{{\bf p}}-k_{\mathrm{r}}\hat{{\bf e}}_{x}$ for pseudo-spin-up atoms and $\hat{{\bf p}}+k_{\mathrm{r}}\hat{{\bf e}}_{x}$ for pseudo-spin-down atoms, respectively. $g_{\sigma\sigma^{\prime}}=4\pi a_{\sigma\sigma^{\prime}}/m$ are interaction strengths for intra- ($\sigma=\sigma^{\prime}$) and interspecies ($\sigma\neq\sigma^{\prime}$), and $a_{\sigma\sigma^{\prime}}$ are the corresponding $s$-wave scattering lengths. $k_{\mathrm{r}}\hat{{\bf e}}_{x}$ is the recoil momentum of the Raman lasers along the $x$ axis, with a recoil energy $E_{\mathrm{r}}=k_{\mathrm{r}}^{2}/(2m)$. The detuning of the Raman lasers is assumed to be zero $\delta=0$, and the Rabi frequency $\Omega$ can be flexibly tuned, in accord with the recent experiments~\cite{ji2014experimental,ji2015softening}. Previous works have shown that the difference between intra- and interspecies interactions is essential to the emergence of the spin-mixed stripe phase~\cite{li2012quantum,martone2014approach}. Hence, we will set $g_{_{\uparrow\uparrow}}=g_{_{\downarrow\downarrow}}=g>g_{_{\uparrow\downarrow}}$ in the following investigations on the stripe phase.

\subsection{The Gross-Pitaevskii equation and Bogoliubov theory}

In this work, we employ the quasiparticle formalism at the Bogoliubov level to describe a weakly interacting dilute Bose gas with SOC at zero temperature~\cite{griffin1996conserving,dodd1998collective,buljan2005incoherent,dalfovo1999theory,Pitaevskii2003Book}. Following the standard procedure~\cite{chen2015collective,chen2017quantum}, the Bose field operator $\hat{\Phi}_{\sigma}({\bf r},t)$ ($\sigma=\uparrow,\downarrow$) can be rewritten as a combination of the condensate wave function $\phi_{\sigma}$ and the noncondensate fluctuation operator $\hat{\eta}_{\sigma}$ as
\begin{equation}
\hat{\Phi}_{\sigma}({\bf r},t)=\phi_{\sigma}({\bf r},t)+\hat{\eta}_{\sigma}({\bf r},t).\label{eq:newBosefield}
\end{equation}
Using a Bogoliubov transformation, the fluctuation operator $\hat{\eta}_{\sigma}({\bf r},t)$ and its conjugate can be expanded as
\begin{subequations} \label{eq:eta}
\begin{eqnarray}
\hat{\eta}_{\sigma} &=& \underset{j}{\sum}\left[u_{j\sigma}({\bf r})e^{-i\varepsilon_{j}t}\hat{a}_{j}+v_{j\sigma}^{*}({\bf r})e^{i\varepsilon_{j}t}\hat{a}_{j}^{\dagger}\right],\\
\hat{\eta}_{\sigma}^{\dagger} &=& \underset{j}{\sum}\left[u_{j\sigma}^{*}({\bf r})e^{i\varepsilon_{j}t}\hat{a}_{j}^{\dagger}+v_{j\sigma}({\bf r})e^{-i\varepsilon_{j}t}\hat{a}_{j}\right],
\end{eqnarray}
\end{subequations}
in terms of the quasiparticle amplitudes $u(u^{*})$, $v(v^{*})$ and the quasiparticle frequency $\varepsilon_{j}$. Here, $j\equiv({\bf q},\tau)$ is the index of the quasiparticle energy level, with the quasimomentum ${\bf q}$ and the branch index $\tau$. $\hat{a}^{\dagger}$ ($\hat{a}$) are respectively the creation (annihilation) operators for quasiparticles, satisfying the bosonic commutation relations: 
\begin{equation}
\left[\hat{a}_{i},\hat{a}^{\dagger}_{j}\right]=\delta_{ij},~\left[\hat{a}_{i}^{\dagger},\hat{a}_{j}^{\dagger}\right]=\left[\hat{a}_{i},\hat{a}_{j}\right]=0.
\end{equation}
After substituting Eq.~\eqref{eq:newBosefield} and Eqs.~\eqref{eq:eta} into the equations of motion 
\begin{equation}
i\partial_{t}\hat{\Phi}_{\sigma}({\bf r},t)=\left[\hat{\Phi}_{\sigma},\hat{H}\right],
\end{equation}
and applying the mean-field decoupling of the cubic terms in $\hat{\Phi}$ and $\hat{\Phi}^\dagger$~\cite{griffin1996conserving}, we obtain two coupled equations as in Ref.~\cite{chen2017quantum}.

The first equation is the modified Gross-Pitaevskii (GP) equation for the condensate,
\begin{equation}
\left[\mathcal{H}_{\mathrm{s}}(\hat{{\bf p}})+\mathrm{diag}(\mathcal{L}_{\uparrow},\mathcal{L}_{\downarrow})\right]\phi=\mu\phi,\label{eq:gp}
\end{equation}
where we have introduced the spinor $\phi\equiv(\phi_{\uparrow},\phi_{\downarrow})^{T}$, the chemical potential $\mu$, and the diagonal element ($\sigma\neq\bar{\sigma}$)
\begin{equation}
\mathcal{L}_{\sigma}\equiv gn_{\sigma}+g_{_{\uparrow\downarrow}}n_{\bar{\sigma}}.
\end{equation}
The second equation is the coupled Bogoliubov equation for quasiparticles,
\begin{subequations} \label{eq:bogoliubov}
\begin{eqnarray}
\left[\mathcal{H}_{\mathrm{s}}(\hat{{\bf p}})-\mu+\mathcal{A}_\uparrow\right]U_j +\mathcal{B}V_j&=&\varepsilon_jU_j, \\
-\mathcal{B}U^*_j -\left[\mathcal{H}_{\mathrm{s}}(\hat{{\bf p}})-\mu+\mathcal{A}_\downarrow\right]V^*_j&=&\varepsilon_jV^*_j,
\end{eqnarray}
\end{subequations}
where $U_{j}\equiv(u_{j\uparrow},u_{j\downarrow})^{T}$, $V_{j}\equiv(v_{j\uparrow},v_{j\downarrow})^{T}$, and 
\begin{equation}
\mathcal{A}_{\sigma}\equiv\left[\begin{array}{cc}
2gn_{\sigma}+g_{_{\uparrow\downarrow}}n_{\bar{\sigma}} & g_{_{\uparrow\downarrow}}\phi_{\sigma}\phi_{\bar{\sigma}}\\
g_{_{\uparrow\downarrow}}\phi_{\bar{\sigma}}\phi_{\sigma} & 2gn_{\bar{\sigma}}+g_{_{\uparrow\downarrow}}n_{\sigma}
\end{array}\right],
\end{equation}
\begin{equation}
\mathcal{B}\equiv\left[\begin{array}{cc}
g\phi_{\uparrow}^{2} & g_{_{\uparrow\downarrow}}\phi_{\uparrow}\phi_{\downarrow}\\
g_{_{\uparrow\downarrow}}\phi_{\uparrow}\phi_{\downarrow} & g\phi_{\downarrow}^{2}
\end{array}\right].
\end{equation}

After solving the GP and Bogoliubov equations, Eqs.~\eqref{eq:gp} and \eqref{eq:bogoliubov}, one obtains straightforwardly the ground-state wave function $\phi({\bf r})$ and the Bogoliubov excitation spectrum $\varepsilon_{j}$, as a function of Rabi frequency $\Omega$. At zero temperature, the total energy of the system can be written in terms of $\phi({\bf r})$ as~\cite{li2012quantum}
\begin{equation}
\begin{aligned}E= & \int d^{3}{\bf r}\left[\left(\phi_{\uparrow}^{\dagger}({\bf r}),\phi_{\downarrow}^{\dagger}({\bf r})\right)\mathcal{H}_{\mathrm{s}}(\hat{{\bf p}})\left(\begin{array}{c}
\phi_{\uparrow}({\bf r})\\
\phi_{\downarrow}({\bf r})
\end{array}\right)\right.\\
 & \left.+\frac{1}{2}g\left(|\phi_{\uparrow}({\bf r})|^{4}+|\phi_{\downarrow}({\bf r})|^{4}\right)+g_{\uparrow\downarrow}|\phi_{\uparrow}({\bf r})|^{2}|\phi_{\downarrow}({\bf r})|^{2}\right].
\end{aligned}
\label{eq:mean-field-energy}
\end{equation}
 In the above derivations, the anomalous densities (i.e., $\langle\hat{\eta}^{\dagger}\hat{\eta}^{\dagger}\rangle$ and $\langle\hat{\eta}\hat{\eta}\rangle$) are omitted, as we use the Popov approximation to ensure a gapless spectrum~\cite{griffin1996conserving,chen2015collective}. In other words, we set the density of the condensate to be the total density, $n_{c}=\bar{n}=N/V$. In addition, the thermal density $\langle\hat{\eta}_{\sigma}^{\dagger}\hat{\eta}_{\sigma}\rangle$ and the spin-flip term $\langle\hat{\eta}_{\sigma}^{\dagger}\hat{\eta}_{\bar{\sigma}}\rangle$ vanish at zero temperature and are therefore not taken into account in our calculations. Nevertheless, the condensate is still depleted by a small fraction of the total density, even at zero temperature, due to quantum fluctuations. This is the so-called quantum depletion,
\begin{equation}
n_{\mathrm{qd}}=\frac{1}{V}\sum_{{\bf q},\tau,\sigma}|v_{\sigma{\bf q}}^{(\tau)}|^{2},\label{eq:quantum_depletion}
\end{equation}
involving typically about $1\%$ of the total density (see next section). The quantum depletion will be explored thoroughly in the next section. Quantum fluctuations also lead to the well-known Lee-Huang-Yang (LHY) correction to the total energy, beyond the mean-field approximation. In our calculations, for the self-consistency of the theory, we do not include the LHY correction to the energy, which is at the order of the square root of the gas parameter $(\bar{n}a^3)^{1/2}$~\cite{zheng2013properties}. Otherwise, the energy correction at the same order due to the anomalous densities may have to be taken into account on an equal footing, which is clearly beyond the scope of our work.

\subsection{The plane-wave ansatz}

The magnetic plane-wave phase and the nonmagnetic zero-momentum phase have been extensively investigated in the previous works~\cite{martone2012anisotropic,zheng2013properties,zhang2016superfluid,chen2017quantum} by using a \textit{plane-wave ansatz} at the momentum $P_{x}$ in Eq.~\eqref{eq:newBosefield}:
\begin{equation}
\phi({\bf r})=\sqrt{\bar{n}}\left(\begin{array}{c}
\cos\theta\\
-\sin\theta
\end{array}\right)e^{iP_{x}x}.\label{eq:plane-wave}
\end{equation}
Here, $\bar{n}=N/V$ is the uniform average density, and the variational angle $\theta$ in the range $[0,\pi/4]$ weighs the spin components of the condensate. In free space, the quasiparticle amplitudes $u_{j\sigma}({\bf r})$, $v_{j\sigma}({\bf r})$ with index $j=({\bf q},\tau)$ and spin component $\sigma$ can be expanded as $u_{j\sigma}({\bf r})=u_{{\bf q}\sigma}^{\tau}e^{i{\bf qr}}$ and $v_{j\sigma}({\bf r})=v_{{\bf q}\sigma}^{\tau}e^{i{\bf qr}}$, where the normalization condition for each branch of two physical solutions ($\tau=\pm$) is given by,
\begin{equation}
\sum_{{\bf q}\sigma}(|u_{{\bf q}\sigma}^{\tau}|^{2}-|v_{{\bf q}\sigma}^{\tau}|^{2})=1.
\end{equation}
In this case, the ground-state energy per particle in Eq.~\eqref{eq:mean-field-energy}
becomes~\cite{li2012quantum,chen2017quantum} 
\begin{equation}  \label{eq:PW-energy}
\begin{aligned}\frac{E^{(\mathrm{PW})}}{N}= & \frac{P_{x}^{2}+k_{\mathrm{r}}^{2}-2P_{x}k_{\mathrm{r}}\cos{2\theta}}{2m}-\frac{1}{2}\Omega\sin{2\theta}\\
 & +\frac{2g\bar{n}-(g-g_{_{\uparrow\downarrow}})\bar{n}\sin^{2}{2\theta}}{4}.
\end{aligned}
\end{equation}
The minimization of the energy gives rise to two solutions: the plane-wave phase where the condensates occur at the momentum $P_{x}=\pm k_{\mathrm{r}}\sqrt{1-\Omega^{2}/[4E_{\mathrm{r}}-(g-g_{\uparrow\downarrow})\bar{n}]^{2}}$ for $\Omega\leq4E_{\mathrm{r}}-(g-g_{\uparrow\downarrow})\bar{n}$ and the zero-momentum phase with zero momentum $P_{x}=0$ for $\Omega>4E_{\mathrm{r}}-(g-g_{\uparrow\downarrow})\bar{n}$~\cite{li2012quantum,chen2017quantum}. In the lowest-lying excitation spectrum, a typical feature of the plane-wave phase is the emergence of the roton-maxon structure. The zero-momentum phase exhibits only the linear phonon mode~\cite{martone2012anisotropic,zheng2013properties,chen2017quantum}.

\subsection{The stripe ansatz}

Instead of the well-studied plane-wave phase and zero-momentum phase, we are concentrating on the exotic stripe phase in this work, which was recently observed in ultracold atomic systems~\cite{leonard2017supersolid,li2017stripe}. To understand the key properties of the stripe phase, a \emph{first-order} \textit{\emph{stripe ansatz}} is often adopted~\cite{li2012quantum,martone2014approach,ji2014experimental,yu2014equation}:
\begin{equation}
\phi(\mathbf{r})=\sqrt{\frac{\bar{n}}{2}}\left[\left(\begin{array}{c}
\sin\theta\\
-\cos\theta
\end{array}\right)e^{-iP_{x}x}+\left(\begin{array}{c}
\cos\theta\\
-\sin\theta
\end{array}\right)e^{iP_{x}x}\right].\label{eq:1st_stripe}
\end{equation}
This ansatz is an equal superposition of two plane waves with momentum $\pm P_{x}$, in contrast to the single-plane-wave ansatz in Eq.~\eqref{eq:plane-wave}.

By substituting this trial wave function, Eq.~\eqref{eq:1st_stripe}, into the model Hamiltonian and minimizing the ground-state energy per particle, which takes the form~\cite{li2012quantum},
\begin{equation}
\begin{aligned}\frac{E^{(\mathrm{1st})}}{N}= & \frac{P_{x}^{2}+k_{\mathrm{r}}^{2}-2P_{x}k_{\mathrm{r}}\cos{2\theta}}{2m}-\frac{1}{2}\Omega\sin{2\theta}\\
 & +\frac{(g+g_{_{\uparrow\downarrow}})\bar{n}}{4}\left(1+\frac{1}{2}\sin^{2}{2\theta}\right),
\end{aligned}
\label{eq:1stST-energy}
\end{equation}
one can straightforwardly determine the critical Rabi frequency $\Omega$ of three exotic phases in the appropriate interaction regimes (i.e., $G_{2}>0$~\footnote{\label{note1}The condition is necessary for the existence of the exotic stripe phase, where the more strict one is $E_{\mathrm{r}}>2G_{2}+2G_{2}^{2}/G_{1}$ in Ref.~\cite{li2012quantum}.}), which are respectively given by~\cite{li2012quantum}
\begin{equation}
\Omega_{c1}=2\left[(2E_{\mathrm{r}}+G_{1})(2E_{\mathrm{r}}-2G_{2})\frac{2G_{2}}{G_{1}+2G_{2}}\right]^{1/2}\label{eq:omega1}
\end{equation}
for the ST-PW phase transition, and 
\begin{equation}
\Omega_{c2}=4E_{\mathrm{r}}-4G_{2}\label{eq:omega2}
\end{equation}
for the PW-ZM phase transition. Here, the two interaction parameters are $G_{1}=(g+g_{\uparrow\downarrow})\bar{n}/4$ and $G_{2}=(g-g_{\uparrow\downarrow})\bar{n}/4$. It is worth mentioning that, in our previous work~\cite{chen2017quantum}, we have determined the ST-PW transition at zero temperature using the criterion of a vanishing roton energy gap. The determined critical Rabi frequency is in a good agreement with $\Omega_{c1}$, and it agrees even better if we neglect quantum fluctuations in the calculations.

The critical Rabi frequency in Eq. \eqref{eq:omega1} for the ST-PW boundary is accurate at the sufficiently weak interaction strengths (i.e., $G_{1}/E_{\textrm{r}},G_{2}/E_{\textrm{r}}\rightarrow0$). However, when the interactions become stronger, a high-order stripe ansatz with high-order harmonics (i.e., the plane waves with wave vectors $\pm3P_{x}$, $\pm5P_{x}$, etc.) need to be considered~\cite{li2013superstripes}. In this work, we take the following stripe \textit{\emph{ansatz}} that includes high-order terms~\cite{li2013superstripes},
\begin{equation}
\phi(\mathbf{r})=\sqrt{\frac{\bar{n}}{2}}\sum_{\gamma=\pm}\sum_{\alpha=1}^{N_{\mathrm{L}}}\left(\begin{array}{c}
\phi_{\uparrow}^{(\gamma\alpha)}\\
\phi_{\downarrow}^{(\gamma\alpha)}
\end{array}\right)e^{i\gamma(2\alpha-1)P_{x}x},\label{eq:high-order_stripe}
\end{equation}
which possesses a symmetry, $\phi_{\uparrow}^{(\alpha)}=-[\phi_{\downarrow}^{(-\alpha)}]^{*}$, and is periodically repeating in real space. Here, $\alpha$ is the index of the stripe order and is smaller or equal to a cutoff integer $N_{\mathrm{L}}$. After solving the wave function $\phi({\bf r})$, the ground-state energy $E^{(N_{\mathrm{L}})}$ can be numerically calculated using Eq.~\eqref{eq:mean-field-energy}. At $N_{\mathrm{L}}=1$, the energy $E^{(N_{\mathrm{L}}=1)}$ recovers the analytic expression for the first-order stripe ansatz $E^{(\mathrm{1st})}$ in Eq. \eqref{eq:1stST-energy}.

To investigate the low-energy excitations, the Bogoliubov quasiparticle amplitudes $u$, $v$ for the index $j=({\bf q},\tau)$ and spin $\sigma$ can be simply expanded in a Bloch form as~\cite{li2013superstripes}
\begin{subequations} \label{eq:uv_nth}
\begin{eqnarray}
u_{j\sigma}({\bf r})&=&e^{i{\bf qr}}\sum_{\gamma=\pm}\sum_{\beta=1}^{N_\mathrm{M}} u^{(\gamma\beta,\tau)}_{\sigma}e^{i\gamma(2\beta-1)P_xx}, \\
v_{j\sigma}({\bf r})&=&e^{i{\bf qr}}\sum_{\gamma=\pm}\sum_{\beta=1}^{N_\mathrm{M}} v^{(\gamma\beta,\tau)}_{\sigma}e^{i\gamma(2\beta-1)P_xx}, 
\end{eqnarray}
\end{subequations}
where ${\bf q}$ is the quasimomentum and $\beta$ is the expansion order index and is smaller than or equal to the cutoff integer $N_{\mathrm{M}}$. We substitute Eqs. \eqref{eq:uv_nth} into the Bogoliubov equations, Eqs. \eqref{eq:bogoliubov}, to determine the expansion coefficients $u_{\sigma}^{(\gamma\beta,\tau)}$ and $v_{\sigma}^{(\gamma\beta,\tau)}$.

In the recent experiments~\cite{lin2011spin,ji2014experimental,ji2015softening}, $^{87}$Rb atoms are used. The typical interaction energy is $gn=0.38E_{\mathrm{r}}$ with the peak density $n=0.46k_{\mathrm{r}}^{3}$ in harmonic traps, and the ratio between the interspecies interaction and intraspecies interaction is very close to unity, i.e., $g_{_{\uparrow\downarrow}}/g=100.99/101.20$~\cite{ji2015softening}. With these parameters, the two critical Rabi frequencies are respectively given by $\Omega_{c1}=0.2E_{\mathrm{r}}$ and $\Omega_{c2}=4.0E_{\mathrm{r}}$ [see Eqs.~\eqref{eq:omega1} and~\eqref{eq:omega2}], characterizing the first-order ST-PW and second-order PW-ZM phase transitions at zero temperature. The stripe phase is energetically favored at only a small region $\Omega\leq0.2E_{\mathrm{r}}$ of the Rabi frequency. The contrast in the stripe density is not large enough to be resolved in the laboratories~\cite{lin2011spin,ji2014experimental,martone2014approach}. In our calculations, we will consider a relatively large ratio of inter- to intraspecies interaction strengths (i.e., large $G_{1}$ and $G_{2}$), in order to enlarge the window for the stripe phase in the phase diagram.

\subsection{The phase-twist method\label{sec:phase-twist}}

Microscopically, by imposing a phase twist ${\bf Q}=Q_{x}\hat{\bf e}_x+Q_{y}\hat{\bf e}_y+Q_{z}\hat{\bf e}_z$, i.e., a supercurrent, on the order parameter 
\begin{equation}
\phi({\bf r})\rightarrow e^{i{\bf Q\cdot r}}\phi({\bf r}), 
\end{equation}
the superfluid will flow with a velocity ${\bf v}_{\mathrm{s}}=\hbar{\bf Q}/m$. In the limit ${\bf Q}\to0$, the variation of free energy $\Delta\mathcal{F}({\bf Q})\equiv\mathcal{F}({\bf Q})-\mathcal{F}(0)$ is approximately given by the extra kinetic energy of the imposed supercurrent, 
\begin{equation}
\begin{aligned}
\Delta\mathcal{F}({\bf Q})\approx & \sum_{i,j=x,y,z}\frac{Q_iQ_j}{2}\lim_{Q_{_{i,j}}\to0}\frac{d^{2}\mathcal{F}({\bf Q})}{dQ_idQ_j}\\
 &\equiv\sum_{i,j}\frac{1}{2}n^{(ij)}_{\mathrm{s}}mv^{(i)}_{\mathrm{s}}v^{(j)}_{\mathrm{s}}V. 
\end{aligned}
\end{equation}
Therefore, the ratio of the tensor element $n^{(ij)}_{\mathrm{s}}$ of the superfluid density over the total density $\bar{n}=N/V$ can be expressed by~\cite{fisher1973helicity,taylor2006pairing,he2018realizing}
\begin{equation} \label{eq:ns_fraction}
\frac{n^{(ij)}_{\mathrm{s}}}{\bar{n}}\equiv\frac{m}{N}\lim_{Q_{_{i,j}}\to0}\frac{d^{2}\mathcal{F}({\bf Q})}{dQ_idQ_j},~i,j=x,y,z.
\end{equation}

In the presence of SOC in $x$ axis, the superfluid density can be written by a tensor form~\cite{zhang2016superfluid}
\begin{equation}
    \hat{n}_{\mathrm{s}}=n^{(x)}_{\mathrm{s}}\hat{\bf e}_x\hat{\bf e}_x+n^{(\perp)}_{\mathrm{s}}(\hat{\bf e}_y\hat{\bf e}_y+\hat{\bf e}_z\hat{\bf e}_z),
\end{equation} 
where the tensor elements at $i\neq j$ vanish due to the reflection symmetry of the Hamiltonian, and $n^{(i=x,\perp)}_{\mathrm{s}}$ indicates the superfluid component along the $x$ direction or in the perpendicular direction, respectively. At zero temperature, without losing the generality, we start with the first-order ansatz in Eq.~\eqref{eq:1st_stripe}, and the corresponding energy per particle $\epsilon(\theta,P_{x})\equiv E/N$ is a function of two variational parameters $(\theta,P_{x})$ (see also Ref.~\cite{li2012quantum}). By fixing the total particle number $N$ and imposing a phase-twist $Q_i$ at the equilibrium $\left(\theta_{0},P_{0}\to\theta(Q_i),P_{0}(Q_i)\right)$~\footnote{In this work, the phase twist is along the direction of SOC, i.e., the $x$ direction, or in the perpendicular $y-z$ plane}, after some straightforward derivations on Eq.~\eqref{eq:ns_fraction}, the fraction of superfluid component can be explicitly expressed by~\cite{he2018realizing}
\begin{equation} \label{eq:ns_SOC}
\frac{n^{(i=x,\perp)}_{\mathrm{s}}}{\bar{n}}=\frac{m}{N}\left[\frac{\partial^{2}\mathcal{F}}{\partial Q_i^{2}}-\left(\frac{\partial^{2}\mathcal{F}}{\partial\theta\partial Q_i}\right)^{2}/\left(\frac{\partial^{2}\mathcal{F}}{\partial\theta^{2}}\right)\right]_{Q_i\to0}.
\end{equation}
This expression is also applicable for the plane-wave and zero-momentum phases.

By substituting Eq.~\eqref{eq:1stST-energy} of the ground-state energy for the stripe phase and Eq.~\eqref{eq:PW-energy} for the plane-wave and zero-momentum phases into Eq.~\eqref{eq:ns_SOC}, we obtain the analytic fraction of the superfluid component in the respective regimes as
\begin{eqnarray}
\left(\frac{n^{(x)}_{\mathrm{s}}}{\bar{n}}\right)_{\textrm{ST}} & = & 1-\frac{2E_{\mathrm{r}}}{(2E_{\mathrm{r}}+G_{1})(4E_{\mathrm{r}}+2G_{1})^{2}/\Omega^{2}-G_{1}},\label{eq:ns_ST}\\
\left(\frac{n^{(x)}_{\mathrm{s}}}{\bar{n}}\right)_{\textrm{PW}} & = & 1-\frac{E_{\mathrm{r}}}{(E_{\mathrm{r}}-G_{2})\Omega_{c2}^{2}/\Omega^{2}+G_{2}},\label{eq:ns_PW}\\
\left(\frac{n^{(x)}_{\mathrm{s}}}{\bar{n}}\right)_{\textrm{ZM}} & = & 1-\frac{4E_{\mathrm{r}}}{\Omega+4G_{2}},\label{eq:ns_ZM}
\end{eqnarray}
in the direction of SOC, i.e., $x$ axis, and
\begin{equation} \label{eq:ns_yz}
\left(\frac{n^{(\perp)}_{\mathrm{s}}}{\bar{n}}\right)_{\textrm{ST}} =\left(\frac{n^{(\perp)}_{\mathrm{s}}}{\bar{n}}\right)_{\textrm{PW}}=\left(\frac{n^{(\perp)}_{\mathrm{s}}}{\bar{n}}\right)_{\textrm{ZM}} =  1
\end{equation}
in the perpendicular plane. The expression in Eq.~\eqref{eq:ns_ST} should be understood as an approximate result for the superfluid fraction in the stripe phase along SOC direction. It can be improved by taking high-order harmonics in the stripe ansatz. The next two expressions, Eqs. \eqref{eq:ns_PW} and \eqref{eq:ns_ZM}, were first obtained in Ref.~\cite{zhang2016superfluid}. It is worth noting that the variational parameters $(\theta,P_{x})$ are independent of the perpendicular twist $Q_{\perp}$ in Eq.~\eqref{eq:ns_SOC}, giving rise to the unaffected superfluid fraction $n^{(\perp)}_{\mathrm{s}}/\bar{n}=1$ in the perpendicular direction, the same as in a usual Bose gas~\cite{zhang2016superfluid}.

\section{Results and Discussions\label{sec:results}}

We are now ready to perform numerical calculations. We take the cutoffs $N_{\mathrm{L}}=N_{\mathrm{M}}\geq14$, to ensure that our results are cutoff independent (see Appendix~\ref{appendix} for the check on the cutoff dependence).

\subsection{Density profile and Bogoliubov excitation spectrum\label{sec:density}}

In this section, we study the density profile and corresponding excitation spectrum of the stripe phase. At zero temperature, we assume the typical interaction energies $G_{1}=0.5E_{\mathrm{r}}$ and $G_{2}=0.1E_{\mathrm{r}}$ with the average density $\bar{n}=1.0k_{\mathrm{r}}^{3}$, which give rise to the critical Rabi frequencies, $\Omega_{c1}=2.27E_{\mathrm{r}}$ and $\Omega_{c2}=3.60E_{\mathrm{r}}$ in Eqs.~\eqref{eq:omega1} and~\eqref{eq:omega2}, within the first-order stripe ansatz.

\begin{figure}
\begin{centering}
\includegraphics[width=0.48\textwidth]{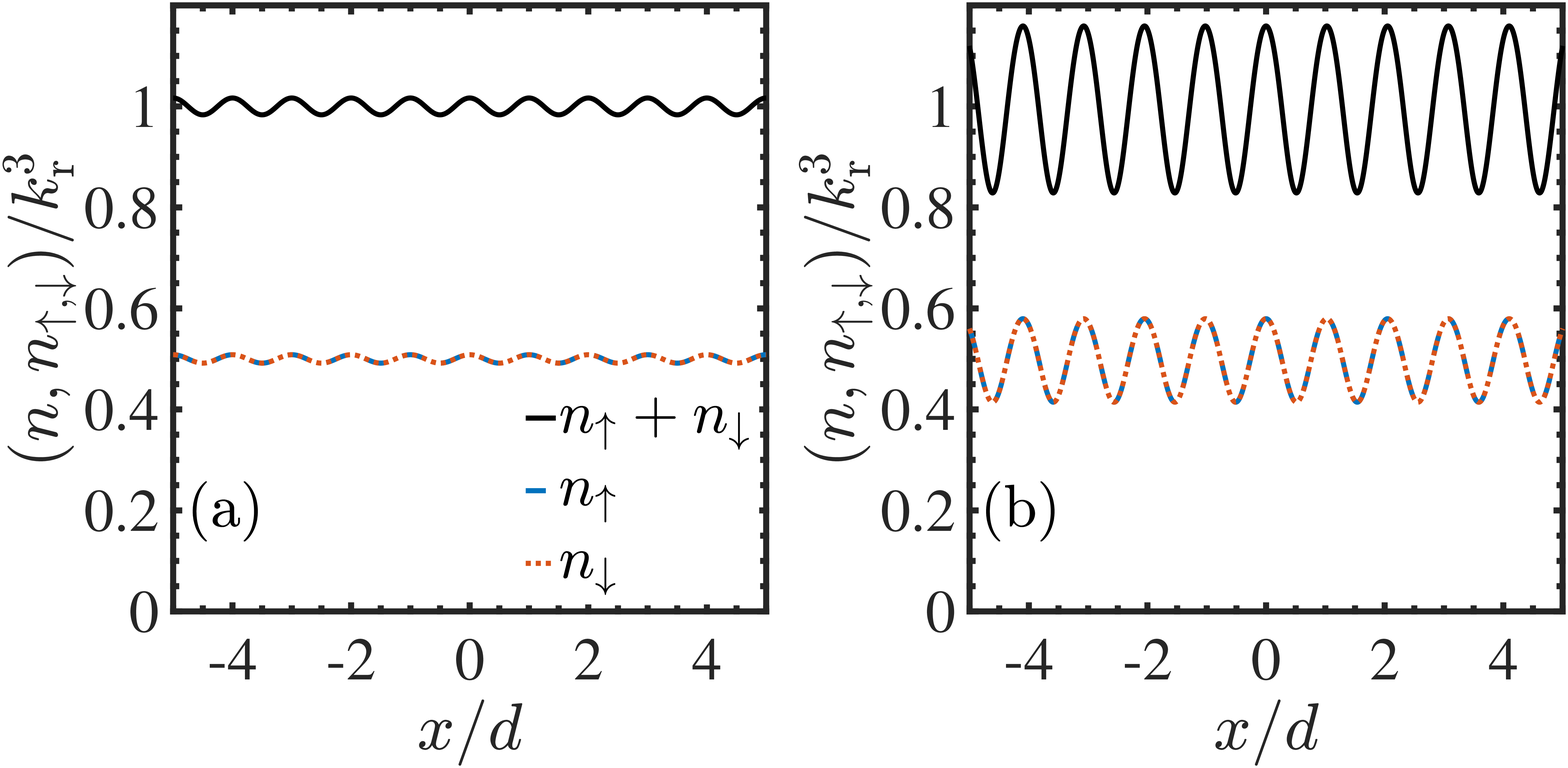}
\par\end{centering}
\begin{centering}
\includegraphics[width=0.48\textwidth]{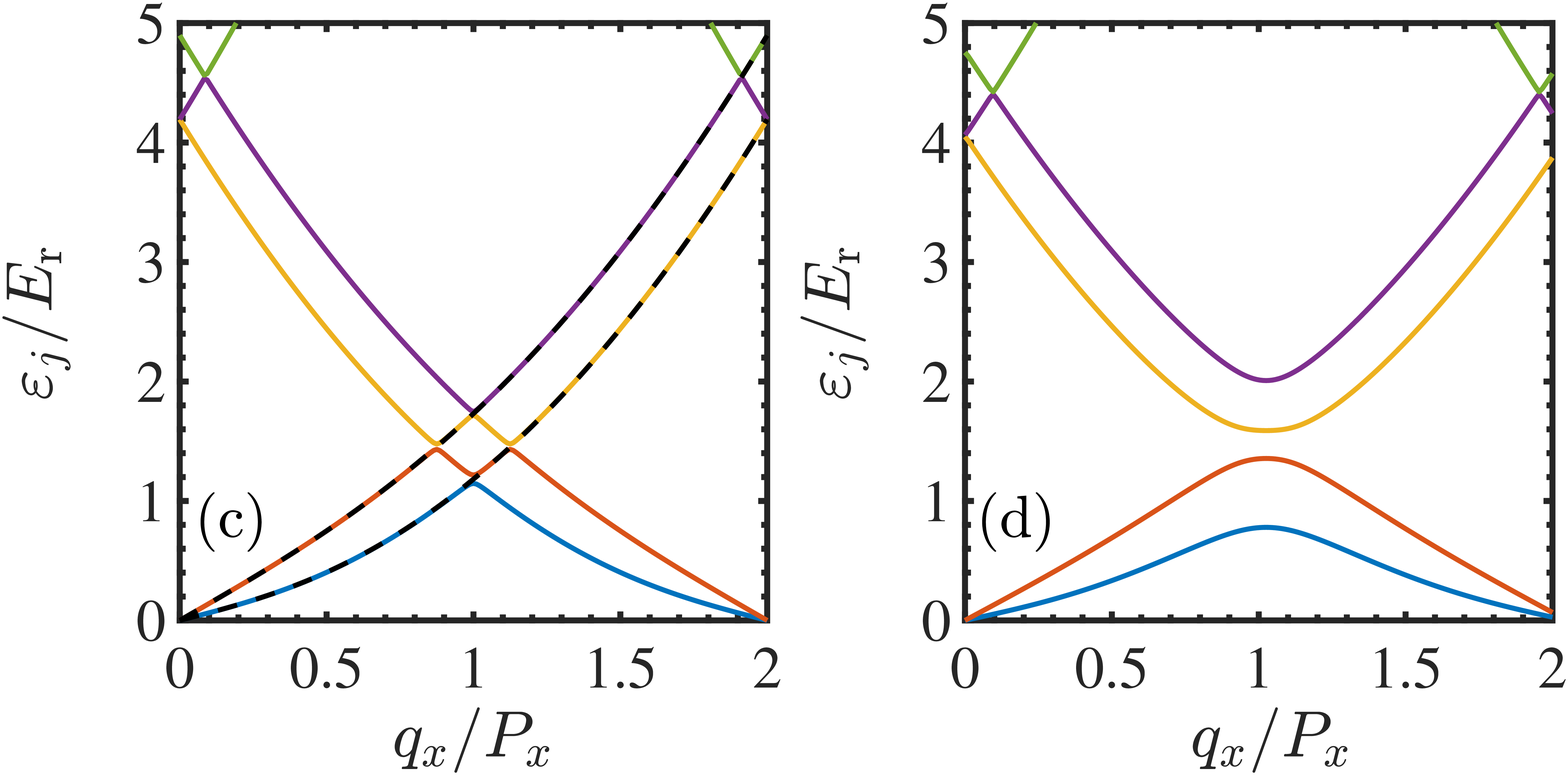}
\par\end{centering}
\centering{}\caption{(Upper panel) The high-order stripe density profile $n$ for spin-up atoms, spin-down atoms, and total atoms along the SOC direction at two Rabi frequencies $\Omega=0.1E_{\mathrm{r}}$ (a) and $\Omega=1.0E_{\mathrm{r}}$ (b). (Lower panel) The corresponding excitation spectrum $\varepsilon_{j}$ for the lowest five branches. Here, we take $G_{1}=0.5E_{\mathrm{r}}$ and $G_{2}=0.1E_{\mathrm{r}}$. $d=\pi/P_{x}$ is the spatial periodicity of stripes. The two dashed lines in (c) show the phonon dispersion of a conventional two-component Bose gas in the limit of $\Omega=0$.}
\label{fig1}
\end{figure}

In Fig.~\ref{fig1}, by setting two different Rabi frequencies $\Omega=0.1E_{\mathrm{r}}$ and $1.0E_{\mathrm{r}}$, we present the respective density distributions and their lowest excitation branches in different colors. In contrast to the plane-wave and zero-momentum phases, the density of the condensate is modulated by the SOC strength in the stripe regime with a spatially periodic order. The total density contrast of the stripe can be estimated using the first-order stripe ansatz and is given by~\cite{li2012quantum,martone2014approach}
\begin{equation}
\mathcal{C}\equiv\frac{n_{\mathrm{max}}-n_{\mathrm{min}}}{n_{\mathrm{max}}+n_{\mathrm{min}}}=\frac{\Omega}{2(2E_{\mathrm{r}}+G_{1})}.
\end{equation}
At $\Omega=0.1E_{\mathrm{r}}$ and $1.0E_{\mathrm{r}}$, the modulation amplitude $\mathcal{C}$ is about $0.02$ and $0.2$ of the total average density $\bar{n}$, respectively. These estimations agree well with the high-order density profiles illustrated in Figs.~\ref{fig1}(a) and \ref{fig1}(b).

\begin{figure}[t]
\centering{}\includegraphics[width=0.48\textwidth]{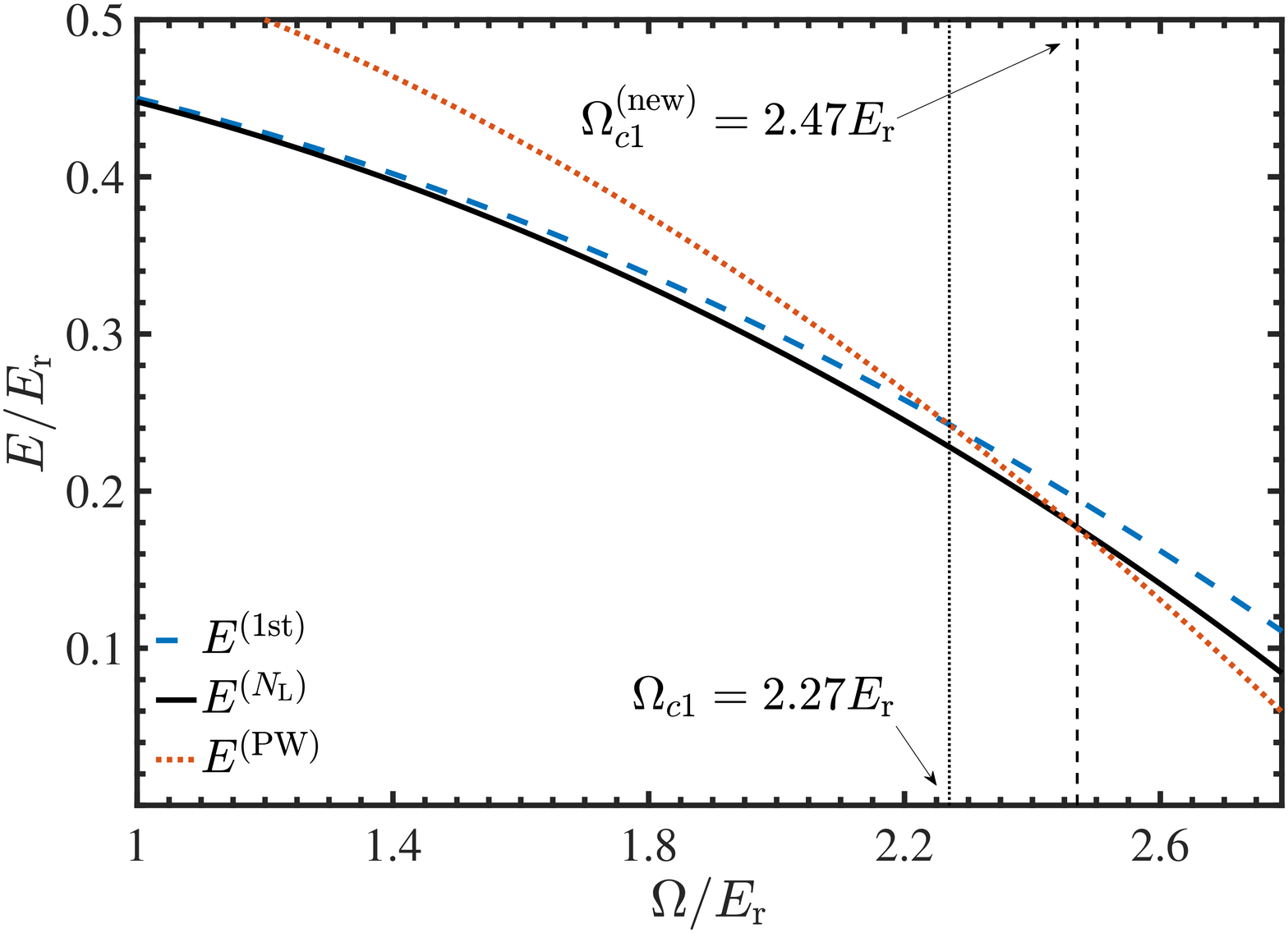} \caption{The ground-state energy as a function of the Rabi frequency in the first-order (dashed-blue), high-order (solid-black) stripe ansatz and the plane-wave ansatz (dotted-red). The interaction parameters are the same as in Fig.~\ref{fig1}.}
\label{fig2} 
\end{figure}

The previous investigations have shown that the lowest-lying excitation spectrum in the plane-wave phase exhibits an intriguing roton-maxon structure due to the degenerate double-minimum in the single-particle dispersion~\cite{martone2012anisotropic,zheng2013properties,mivehvar2015enhanced,chen2017quantum}. As the Rabi frequency decreases towards the ST-PW transition, the roton structure becomes much more clear and the roton energy gap is gradually approaching zero, indicating a critical ST-PW Rabi frequency~\cite{chen2017quantum}. However, in the stripe phase, the density modulation spontaneously breaks the spatial translational symmetry, giving rise to infinite gapped branches as a function of the quasimomentum $q_{x}\in[0,2P_{x}]$. The lowest two branches are gapless~\cite{li2013superstripes}, as indicated by two linear phonon modes, i.e., the red and blue curves in Figs.~\ref{fig1}(c) and \ref{fig1}(d). As we decrease the Rabi frequency towards the limit $\Omega\rightarrow0$, the gap between different excitation branches vanishes and one recovers the Bogoliubov excitation spectrum~\cite{abad2013study}
\begin{equation}
\omega_{\pm}({\bf k})=\sqrt{\frac{{\bf k}^{2}}{2m}\left[\frac{{\bf k}^{2}}{2m}+(g\pm g_{_{\uparrow\downarrow}})\bar{n}\right]},
\end{equation}
anticipated for a conventional two-component Bose gas {[}see the two dashed black curves in Fig.~\ref{fig1}(c){]}.

\subsection{Critical Rabi frequency for the ST-PW phase transition}

\begin{figure}[t]
\centering{}\includegraphics[width=0.48\textwidth]{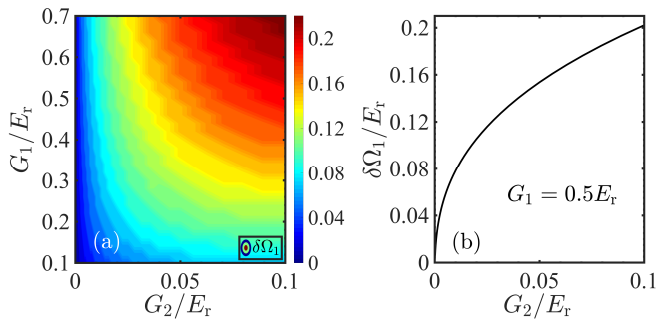}\caption{(a) Contour plot of the shift $\delta\Omega_{1}=\Omega_{c1}^{\mathrm{(new)}}-\Omega_{c1}$ for the ST-PW transition as functions of the interaction energy strengths $G_{1}$ and $G_{2}$. (b) The dependence of $\delta\Omega_{1}$ on $G_{2}$ at $G_{1}=0.5E_{\mathrm{r}}$.}
\label{fig3} 
\end{figure}

The two critical Rabi frequencies, $\Omega_{c1}$ and $\Omega_{c2}$, are determined by comparing the total energy $E^{(\mathrm{1st})}$ of the first-order stripe and $E^{(\mathrm{PW})}$ of the plane-wave ansatz [see, e.g., Eqs.~\eqref{eq:1stST-energy} and~\eqref{eq:PW-energy}]. By taking into account the high-order harmonics in Eq.~\eqref{eq:high-order_stripe}, the stripe phase may become energetically more favorable, and as a consequence, the first-order ST-PW transition point may shift to a relatively larger Rabi frequency. This is indeed confirmed in Fig.~\ref{fig2}. At the same interaction parameters, the ground-state energy of the stripe phase becomes lower with the inclusion of high-order harmonics, i.e., $E^{(N_{\mathrm{L}})}\leq E^{(\mathrm{1st})}$. The window for the stripe phase is therefore enlarged, with a larger critical Rabi frequency $\Omega_{c1}^{\mathrm{(new)}}=2.47E_{\mathrm{r}}>\Omega_{c1}=2.27E_{\textrm{r}}$ for the first-order ST-PW transition.

In Fig.~\ref{fig3}(a), we show the dependence of the relative shift of the ST-PW transition, $\delta\Omega_{1}\equiv\Omega_{c1}^{\mathrm{(new)}}-\Omega_{c1}$, on the interaction parameters $G_{1}$ (the vertical axis) and $G_{2}$ (the horizontal axis). Figure~\ref{fig3}(b) reports the shift as a function of $G_{2}$ at $G_{1}=0.5E_{\mathrm{r}}$.

\begin{figure*}
\centering{}\includegraphics[width=0.96\textwidth]{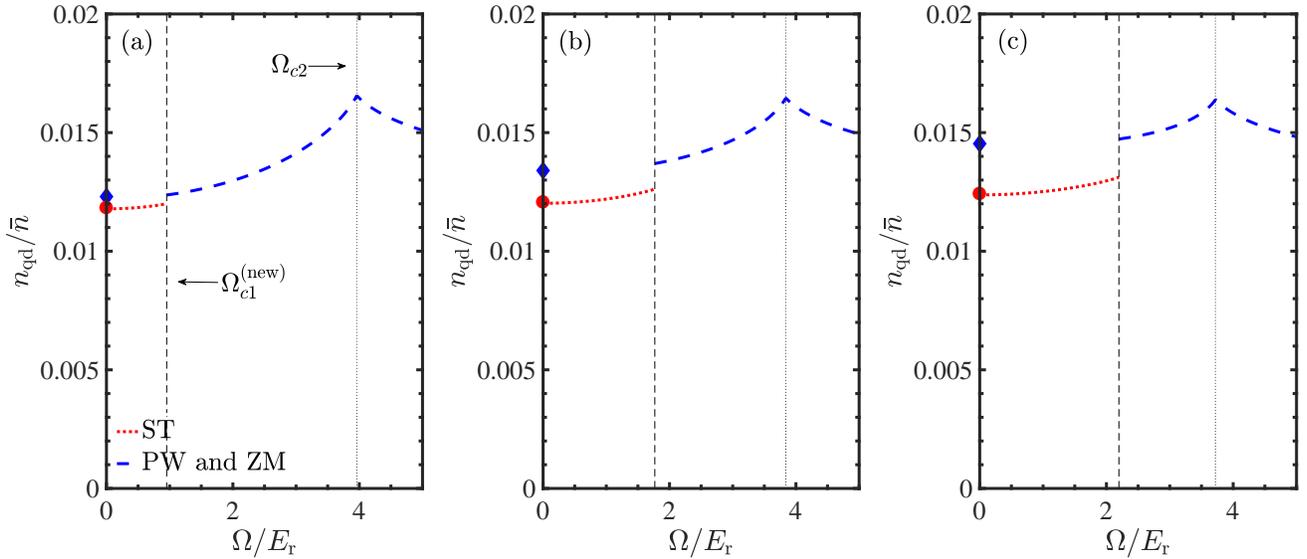}\caption{Quantum depletion $n_{\mathrm{qd}}/\bar{n}$ as a function of the Rabi frequency $\Omega$ in the ST phase (red dotted line) and in the PW and ZM phases (blue dashed line). The blue diamonds show the depletion of a uniform single-component weakly interacting Bose gas with the same interaction parameters, while the red circles give the depletion of a two-component Bose gas. The vertical dashed and dotted curves indicate the critical $\Omega_{c1}^{(\mathrm{new})}$ and $\Omega_{c2}$, respectively. Here, we take the interaction energies $G_{1}=0.5E_{\mathrm{r}}$ and $G_{2}=0.01E_{\textrm{r}}$ (a), $0.04E_{\mathrm{r}}$ (b), and $0.07E_{\mathrm{r}}$ (c).}
\label{fig4} 
\end{figure*}

In general, at sufficiently small $G_{2}$ (i.e., $\ll0.01E_{\mathrm{r}}$), the stripe phase locates at a very small range of the Rabi frequency, where the density is slightly modulated with negligible contrast. We find that the difference $\delta\Omega_{1}$, as shown in Fig.~\ref{fig3}(b), is close to zero. This implies a negligible contribution of the high-order harmonics, and the state of the system can be well described by the dominant first-order stripe trial wave function. When the interaction energy $G_{2}$ becomes relatively larger, the density modulation becomes significant and the difference $\delta\Omega_{1}$ is sizable, as shown by the yellow and red regions in Fig.~\ref{fig3}(a). Meanwhile, as $G_{1}$ increases, $\delta\Omega_{1}$ becomes much more pronounced. The significant shift of the ST-PW phase transition position indicates the crucial role played by the high-order harmonics in Eq.~\eqref{eq:high-order_stripe}. Our results suggest that they have to be accounted for in future theoretical investigations, particularly at relatively large values of interaction parameters.

\subsection{Quantum depletion at zero temperature}

Using the mean-field Bogoliubov theory, it is straightforward to obtain the quantum depletion using Eq.~\eqref{eq:quantum_depletion}. For a single-component Bose gas, the quantum depletion was recently measured~\cite{lopes2017quantum}. In Fig.~\ref{fig4}, we present the $\Omega$ dependence of the quantum depletion at zero temperature across all three phases, at $G_{1}=0.5E_{\mathrm{r}}$ and at three different values of $G_{2}$ (i.e., $0.01E_{\mathrm{r}}$, $0.04E_{\mathrm{r}}$, and $0.07E_{\mathrm{r}}$).

The quantum depletion in the plane-wave and zero-momentum phases has been studied in Ref. \cite{zhang2016superfluid}. We show their behavior in great detail in Figs. \ref{fig4}(a)--\ref{fig4}(c) (i.e., the dashed-blue curves). It is a nonmonotonic function of the Rabi frequency. In the plane-wave phase, the contribution to quantum depletion comes from both phonons and rotons in the lowest-lying excitation spectrum~\cite{ji2015softening}. As one increases $\Omega$, the roton energy gap becomes larger and the phonon mode dominates the contribution. This leads to a maximum in the depletion at the PW-ZM transition $\Omega_{c2}$ (i.e., dotted-black vertical curve)~\cite{zheng2013properties}. As $\Omega$ continues to increase in the zero-momentum phase, the depletion decreases, since the roton contribution to the depletion disappears and the only contribution is from the phonon mode.

The behavior of depletion in the stripe regime (i.e., the dotted-red curves) has not been studied before. It increases slowly and monotonically, as $\Omega$ increases up to the ST-PW transition $\Omega_{c1}^{(\mathrm{new})}$. This can be understood from the smooth softening of the two phonon modes, as illustrated in Figs.~\ref{fig1}(c) and \ref{fig1}(d). It is worth noting that the depletion at $\Omega_{c1}^{(\mathrm{new})}$ experiences a jump due to the first-order character of the transition~\cite{li2012quantum,martone2012anisotropic,chen2017quantum}. The size of the discontinuity becomes significant as we increase $G_{2}$.

\begin{figure*}
\centering{}\includegraphics[width=0.96\textwidth]{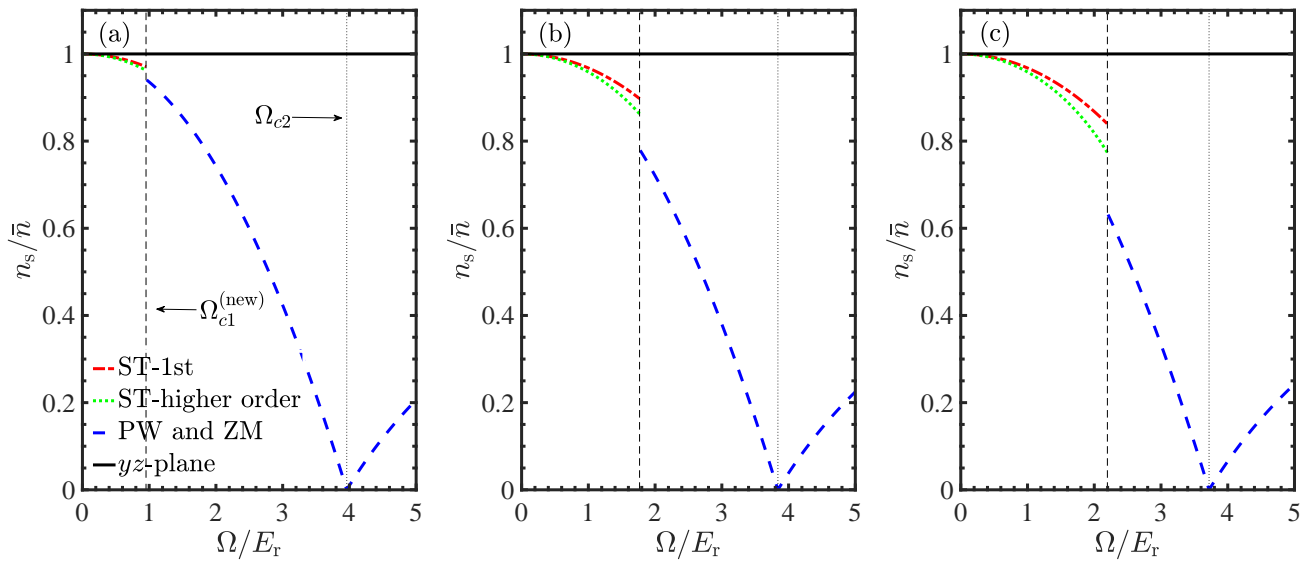}\caption{Superfluid fraction $n^{(x)}_{\mathrm{s}}/\bar{n}$ as a function of the Rabi frequency $\Omega$ in the ST phase (red dashed line -- first-order ansatz; green dotted line -- high-order ansatz), the PW and ZM phases (blue dashed line). The solid-black lines are the component $n^{(\perp)}_{\mathrm{s}}/\bar{n}$ in the perpendicular plane. The two vertical lines indicate the critical Rabi frequency of the phase transition. The interaction parameters are the same as in Fig.~\ref{fig4}.}
\label{fig5} 
\end{figure*}

The behavior of the quantum depletion can be further understood with two analytic results. In the absence of SOC, the quantum depletion of a conventional homogeneous single-component Bose gas is of the order of $\sqrt{\bar{n}a^{3}}$ and can be analytically written as~\cite{pethick2002bose,Pitaevskii2003Book}
\begin{equation}
\frac{n_{\mathrm{qd}}^{(\mathrm{1c})}}{\bar{n}}=\frac{8}{3\sqrt{\pi}}(\bar{n}a^{3})^{1/2}.
\end{equation}
This result can be extended to the two-component case with equal spin
density as 
\begin{eqnarray}
\frac{n_{\mathrm{qd}}^{(\mathrm{2c})}}{\bar{n}} & = & \frac{8}{3\sqrt{\pi}}\left[(\bar{n}a_{+}^{3})^{1/2}+(\bar{n}a_{-}^{3})^{1/2}\right],
\end{eqnarray}
where the scattering length $a_{\pm}=(a\pm a_{\uparrow\downarrow})/2$ and $a\equiv a_{\uparrow\uparrow}=a_{\downarrow\downarrow}$. In Fig.~\ref{fig4}, we have checked that by starting with the plane-wave ansatz in Eq.~\eqref{eq:plane-wave}, towards the limit $\Omega\rightarrow0$, the depletion coincides with the single-component $n_{\mathrm{qd}}^{(\mathrm{1c})}$ (see blue diamonds), as the plane-wave phase tends to be fully spin polarized. Similarly, the depletion predicted using the stripe ansatz in Eq.~\eqref{eq:high-order_stripe} (i.e., equal combination of two spins) recovers the two-component $n_{\mathrm{qd}}^{(\mathrm{2c})}$ (see red circles) in the limit $\Omega\rightarrow0$.

\subsection{Superfluid fraction at zero temperature}

We now turn to discuss the superfluidity of the system in the presence of SOC, which can be characterized by the superfluid density. In Fig.~\ref{fig5}, we report the behavior of the superfluid density as a function of Rabi frequency in three phases at the same interaction energy strengths as in Fig.~\ref{fig4}.

It is apparent from the figure that the superfluid density $n^{(x)}_{\mathrm{s}}$ along the SOC direction in the plane-wave and zero-momentum regimes (i.e., the dashed-blue curves), calculated by using Eqs.~\eqref{eq:ns_PW} and~\eqref{eq:ns_ZM}, exhibits an intriguing behavior. It goes down monotonically in the plane-wave phase, touches zero at the PW-ZM transition $\Omega_{c2}$, and then bounces back in the zero-momentum phase. This behavior exactly recovers the result in Ref.~\cite{zhang2016superfluid}, where the normal density of the system was calculated using the transverse current response function at zero temperature and the nonzero normal density was explained using a sum rule together with a gapped branch in the elementary excitation spectrum~\cite{zhang2016superfluid}.

On the contrary, in the stripe phase the superfluid density can be evaluated using the first-order stripe ansatz. The resulting analytic prediction [see Eq.~\eqref{eq:ns_ST}] is shown by the red dotted-dashed curves in Fig.~\ref{fig5}. Using the same phase-twist method but with the high-order stripe ansatz, we obtain the dotted-green curves in the figure. The difference between the two results (i.e., first-order vs high-order) increases with increasing $\Omega$. The difference also becomes significant if we use a large interaction parameter $G_{2}$. The suppression of the superfluid density at nonzero $\Omega$ may be understood from the softening of the lowest two phonon modes in the low-energy excitation spectrum [see Figs.~\ref{fig1}(c) and \ref{fig1}(d)]. The critical velocity decreases as the phonon modes become soft. This means that physically it is more favorable to create excitations to destroy the superfluidity of the system~\cite{Pitaevskii2003Book}. As anticipated, the superfluid density also exhibits a discontinuity at $\Omega_{c1}^{(\mathrm{new})}$, because of the first-order ST-PW transition.

It is worth noting that in the perpendicular plane, there is no density modulation due to the absence of the spin-orbit coupling. As a result, the superfluid fraction $n^{(\perp)}_{\mathrm{s}}/\bar{n}$ is 100\% (see the solid-black curves), the same as in a conventional spinless Bose gas.

\section{Conclusions and Outlooks\label{sec:summary}}

In conclusion, we have applied the mean-field Gross-Pitaevskii equation and Bogoliubov theory to characterize the stripe phase of a Raman-type spin-orbit-coupled Bose gas at zero temperature. The stripe density of the condensate, which is significantly modulated by spin-orbit coupling, has been calculated, and the corresponding low-energy excitation spectrum has been explored over a large range of the Rabi frequency. By using a high-order stripe ansatz, we have determined in a more accurate way the critical Rabi frequency for the transition between the stripe and plane-wave phases. We have calculated the quantum depletion in all three phases of the system. We have also derived an explicit but approximate expression of the superfluid density within a phase-twist approach. This first-order analytic prediction has been compared with the more accurate numerical result obtained with a stripe ansatz that involves high-order harmonics. Further questions, such as the finite-temperature effect and other physical observables such as the moment of inertia~\cite{stringari2017diffused}, remain to be investigated in order to better understand the exotic stripe phase.
\begin{acknowledgments}
Our research was supported through the Australian Research Council's (ARC) Discovery Projects No. FT140100003 and No. DP180102018 (X.J.L.), and No. FT130100815 and No. DP170104008 (H.H.).
\end{acknowledgments}

\appendix

\begin{figure}[b]
\centering{}\includegraphics[width=0.48\textwidth]{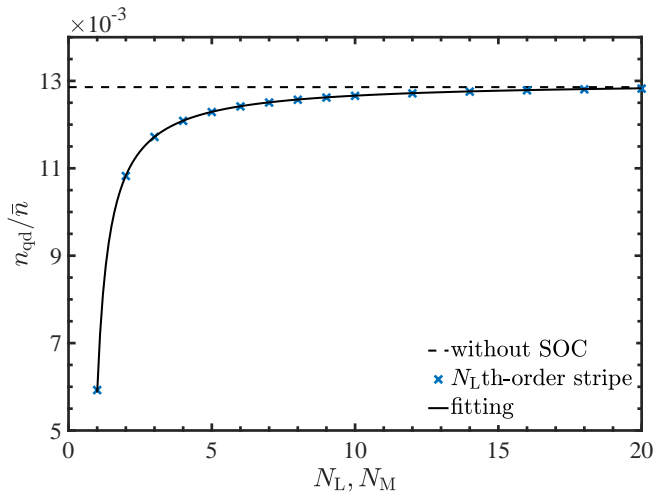} \caption{\label{fig6} (color online). Quantum depletion in the limit of $\Omega\to0$, as a function of $N_{\mathrm{L}}=N_{\textrm{M}}$. The anticipated result of a two-component Bose gas is shown by a dashed line. The interaction parameters are the same as in Fig.~\ref{fig1}.}
\end{figure}

\section{The choice of the cutoffs \texorpdfstring{$N_{\rm L}$}{Lg} and \texorpdfstring{$N_{\rm M}$}{Lg}  \label{appendix}}

In this Appendix, we check the convergence of our numerical calculations with respect to the cutoffs $N_{\mathrm{L}}$ and $N_{\mathrm{M}}$ by calculating the quantum depletion. In principle, $N_{\mathrm{L}}$ and $N_{\mathrm{M}}$ should be infinitely large to include all high-order terms. However, in practice we have to use finite cutoffs to ensure the computational efficiency. In Fig.~\ref{fig6}, we show the depletion (blue crosses) at different sets of cutoffs $N_{\mathrm{L}}$ and $N_{\mathrm{M}}$ in the zero Rabi frequency limit $\Omega\to0$. In this limit, the quantum depletion is analytically known (see the dashed line), as the system reduces to a uniform two-component Bose gas without spin-orbit coupling. At $N_{\mathrm{L}}=N_{\mathrm{M}}=14$, we find that the relative deviation of the calculated depletion is less than $1\%$. Therefore, we use $N_{\mathrm{L}}\geq14$ and $N_{\mathrm{M}}\geq14$ in our numerical calculations.

\bibliographystyle{apsrev4-1}
\bibliography{bibliography.bib}

\providecommand{\noopsort}[1]{}\providecommand{\singleletter}[1]{#1}%
\begin{thebibliography}{64}%
\makeatletter
\providecommand \@ifxundefined [1]{%
 \@ifx{#1\undefined}
}%
\providecommand \@ifnum [1]{%
 \ifnum #1\expandafter \@firstoftwo
 \else \expandafter \@secondoftwo
 \fi
}%
\providecommand \@ifx [1]{%
 \ifx #1\expandafter \@firstoftwo
 \else \expandafter \@secondoftwo
 \fi
}%
\providecommand \natexlab [1]{#1}%
\providecommand \enquote  [1]{``#1''}%
\providecommand \bibnamefont  [1]{#1}%
\providecommand \bibfnamefont [1]{#1}%
\providecommand \citenamefont [1]{#1}%
\providecommand \href@noop [0]{\@secondoftwo}%
\providecommand \href [0]{\begingroup \@sanitize@url \@href}%
\providecommand \@href[1]{\@@startlink{#1}\@@href}%
\providecommand \@@href[1]{\endgroup#1\@@endlink}%
\providecommand \@sanitize@url [0]{\catcode `\\12\catcode `\$12\catcode
  `\&12\catcode `\#12\catcode `\^12\catcode `\_12\catcode `\%12\relax}%
\providecommand \@@startlink[1]{}%
\providecommand \@@endlink[0]{}%
\providecommand \url  [0]{\begingroup\@sanitize@url \@url }%
\providecommand \@url [1]{\endgroup\@href {#1}{\urlprefix }}%
\providecommand \urlprefix  [0]{URL }%
\providecommand \Eprint [0]{\href }%
\providecommand \doibase [0]{http://dx.doi.org/}%
\providecommand \selectlanguage [0]{\@gobble}%
\providecommand \bibinfo  [0]{\@secondoftwo}%
\providecommand \bibfield  [0]{\@secondoftwo}%
\providecommand \translation [1]{[#1]}%
\providecommand \BibitemOpen [0]{}%
\providecommand \bibitemStop [0]{}%
\providecommand \bibitemNoStop [0]{.\EOS\space}%
\providecommand \EOS [0]{\spacefactor3000\relax}%
\providecommand \BibitemShut  [1]{\csname bibitem#1\endcsname}%
\let\auto@bib@innerbib\@empty
\bibitem [{\citenamefont {Galitski}\ and\ \citenamefont
  {Spielman}(2013)}]{Galitski2013}%
  \BibitemOpen
  \bibfield  {author} {\bibinfo {author} {\bibfnamefont {V.}~\bibnamefont
  {Galitski}}\ and\ \bibinfo {author} {\bibfnamefont {I.~B.}\ \bibnamefont
  {Spielman}},\ }\href {\doibase 10.1038/nature11841} {\bibfield  {journal}
  {\bibinfo  {journal} {Nature}\ }\textbf {\bibinfo {volume} {494}},\ \bibinfo
  {pages} {49} (\bibinfo {year} {2013})}\BibitemShut {NoStop}%
\bibitem [{\citenamefont {Zhai}(2015)}]{zhai2015degenerate}%
  \BibitemOpen
  \bibfield  {author} {\bibinfo {author} {\bibfnamefont {H.}~\bibnamefont
  {Zhai}},\ }\href {http://stacks.iop.org/0034-4885/78/i=2/a=026001} {\bibfield
   {journal} {\bibinfo  {journal} {Reports on Progress in Physics}\ }\textbf
  {\bibinfo {volume} {78}},\ \bibinfo {pages} {026001} (\bibinfo {year}
  {2015})}\BibitemShut {NoStop}%
\bibitem [{\citenamefont {Zhu}\ \emph {et~al.}(2006)\citenamefont {Zhu},
  \citenamefont {Fu}, \citenamefont {Wu}, \citenamefont {Zhang},\ and\
  \citenamefont {Duan}}]{zhu2006spin}%
  \BibitemOpen
  \bibfield  {author} {\bibinfo {author} {\bibfnamefont {S.-L.}\ \bibnamefont
  {Zhu}}, \bibinfo {author} {\bibfnamefont {H.}~\bibnamefont {Fu}}, \bibinfo
  {author} {\bibfnamefont {C.-J.}\ \bibnamefont {Wu}}, \bibinfo {author}
  {\bibfnamefont {S.-C.}\ \bibnamefont {Zhang}}, \ and\ \bibinfo {author}
  {\bibfnamefont {L.-M.}\ \bibnamefont {Duan}},\ }\href {\doibase
  10.1103/PhysRevLett.97.240401} {\bibfield  {journal} {\bibinfo  {journal}
  {Phys. Rev. Lett.}\ }\textbf {\bibinfo {volume} {97}},\ \bibinfo {pages}
  {240401} (\bibinfo {year} {2006})}\BibitemShut {NoStop}%
\bibitem [{\citenamefont {Liu}\ \emph {et~al.}(2007)\citenamefont {Liu},
  \citenamefont {Liu}, \citenamefont {Kwek},\ and\ \citenamefont
  {Oh}}]{liu2007optically}%
  \BibitemOpen
  \bibfield  {author} {\bibinfo {author} {\bibfnamefont {X.-J.}\ \bibnamefont
  {Liu}}, \bibinfo {author} {\bibfnamefont {X.}~\bibnamefont {Liu}}, \bibinfo
  {author} {\bibfnamefont {L.~C.}\ \bibnamefont {Kwek}}, \ and\ \bibinfo
  {author} {\bibfnamefont {C.~H.}\ \bibnamefont {Oh}},\ }\href {\doibase
  10.1103/PhysRevLett.98.026602} {\bibfield  {journal} {\bibinfo  {journal}
  {Phys. Rev. Lett.}\ }\textbf {\bibinfo {volume} {98}},\ \bibinfo {pages}
  {026602} (\bibinfo {year} {2007})}\BibitemShut {NoStop}%
\bibitem [{\citenamefont {Beeler}\ \emph {et~al.}(2013)\citenamefont {Beeler},
  \citenamefont {Williams}, \citenamefont {Jimenez-Garcia}, \citenamefont
  {Perry},\ and\ \citenamefont {Spielman}}]{Beeler2013}%
  \BibitemOpen
  \bibfield  {author} {\bibinfo {author} {\bibfnamefont {M.~C.}\ \bibnamefont
  {Beeler}}, \bibinfo {author} {\bibfnamefont {R.~A.}\ \bibnamefont
  {Williams}}, \bibinfo {author} {\bibfnamefont {L.~J.}\ \bibnamefont
  {Jimenez-Garcia}, \bibfnamefont {K.and~LeBlanc}}, \bibinfo {author}
  {\bibfnamefont {A.~R.}\ \bibnamefont {Perry}}, \ and\ \bibinfo {author}
  {\bibfnamefont {I.~B.}\ \bibnamefont {Spielman}},\ }\href
  {http://dx.doi.org/10.1038/nature12185} {\bibfield  {journal} {\bibinfo
  {journal} {Nature}\ }\textbf {\bibinfo {volume} {498}},\ \bibinfo {pages}
  {201} (\bibinfo {year} {2013})}\BibitemShut {NoStop}%
\bibitem [{\citenamefont {Sato}\ \emph {et~al.}(2009)\citenamefont {Sato},
  \citenamefont {Takahashi},\ and\ \citenamefont {Fujimoto}}]{sato2009non}%
  \BibitemOpen
  \bibfield  {author} {\bibinfo {author} {\bibfnamefont {M.}~\bibnamefont
  {Sato}}, \bibinfo {author} {\bibfnamefont {Y.}~\bibnamefont {Takahashi}}, \
  and\ \bibinfo {author} {\bibfnamefont {S.}~\bibnamefont {Fujimoto}},\ }\href
  {\doibase 10.1103/PhysRevLett.103.020401} {\bibfield  {journal} {\bibinfo
  {journal} {Phys. Rev. Lett.}\ }\textbf {\bibinfo {volume} {103}},\ \bibinfo
  {pages} {020401} (\bibinfo {year} {2009})}\BibitemShut {NoStop}%
\bibitem [{\citenamefont {Jiang}\ \emph {et~al.}(2011)\citenamefont {Jiang},
  \citenamefont {Kitagawa}, \citenamefont {Alicea}, \citenamefont {Akhmerov},
  \citenamefont {Pekker}, \citenamefont {Refael}, \citenamefont {Cirac},
  \citenamefont {Demler}, \citenamefont {Lukin},\ and\ \citenamefont
  {Zoller}}]{jiang2011majorana}%
  \BibitemOpen
  \bibfield  {author} {\bibinfo {author} {\bibfnamefont {L.}~\bibnamefont
  {Jiang}}, \bibinfo {author} {\bibfnamefont {T.}~\bibnamefont {Kitagawa}},
  \bibinfo {author} {\bibfnamefont {J.}~\bibnamefont {Alicea}}, \bibinfo
  {author} {\bibfnamefont {A.~R.}\ \bibnamefont {Akhmerov}}, \bibinfo {author}
  {\bibfnamefont {D.}~\bibnamefont {Pekker}}, \bibinfo {author} {\bibfnamefont
  {G.}~\bibnamefont {Refael}}, \bibinfo {author} {\bibfnamefont {J.~I.}\
  \bibnamefont {Cirac}}, \bibinfo {author} {\bibfnamefont {E.}~\bibnamefont
  {Demler}}, \bibinfo {author} {\bibfnamefont {M.~D.}\ \bibnamefont {Lukin}}, \
  and\ \bibinfo {author} {\bibfnamefont {P.}~\bibnamefont {Zoller}},\ }\href
  {\doibase 10.1103/PhysRevLett.106.220402} {\bibfield  {journal} {\bibinfo
  {journal} {Phys. Rev. Lett.}\ }\textbf {\bibinfo {volume} {106}},\ \bibinfo
  {pages} {220402} (\bibinfo {year} {2011})}\BibitemShut {NoStop}%
\bibitem [{\citenamefont {Liu}\ \emph {et~al.}(2012)\citenamefont {Liu},
  \citenamefont {Jiang}, \citenamefont {Pu},\ and\ \citenamefont
  {Hu}}]{liu2012probing}%
  \BibitemOpen
  \bibfield  {author} {\bibinfo {author} {\bibfnamefont {X.-J.}\ \bibnamefont
  {Liu}}, \bibinfo {author} {\bibfnamefont {L.}~\bibnamefont {Jiang}}, \bibinfo
  {author} {\bibfnamefont {H.}~\bibnamefont {Pu}}, \ and\ \bibinfo {author}
  {\bibfnamefont {H.}~\bibnamefont {Hu}},\ }\href {\doibase
  10.1103/PhysRevA.85.021603} {\bibfield  {journal} {\bibinfo  {journal} {Phys.
  Rev. A}\ }\textbf {\bibinfo {volume} {85}},\ \bibinfo {pages} {021603}
  (\bibinfo {year} {2012})}\BibitemShut {NoStop}%
\bibitem [{\citenamefont {Liu}\ and\ \citenamefont
  {Hu}(2012)}]{liu2012topological}%
  \BibitemOpen
  \bibfield  {author} {\bibinfo {author} {\bibfnamefont {X.-J.}\ \bibnamefont
  {Liu}}\ and\ \bibinfo {author} {\bibfnamefont {H.}~\bibnamefont {Hu}},\
  }\href {\doibase 10.1103/PhysRevA.85.033622} {\bibfield  {journal} {\bibinfo
  {journal} {Phys. Rev. A}\ }\textbf {\bibinfo {volume} {85}},\ \bibinfo
  {pages} {033622} (\bibinfo {year} {2012})}\BibitemShut {NoStop}%
\bibitem [{\citenamefont {Liu}\ and\ \citenamefont
  {Hu}(2013)}]{liu2013topological}%
  \BibitemOpen
  \bibfield  {author} {\bibinfo {author} {\bibfnamefont {X.-J.}\ \bibnamefont
  {Liu}}\ and\ \bibinfo {author} {\bibfnamefont {H.}~\bibnamefont {Hu}},\
  }\href {\doibase 10.1103/PhysRevA.88.023622} {\bibfield  {journal} {\bibinfo
  {journal} {Phys. Rev. A}\ }\textbf {\bibinfo {volume} {88}},\ \bibinfo
  {pages} {023622} (\bibinfo {year} {2013})}\BibitemShut {NoStop}%
\bibitem [{\citenamefont {Liu}\ \emph {et~al.}(2014)\citenamefont {Liu},
  \citenamefont {Law},\ and\ \citenamefont {Ng}}]{liu2014realization}%
  \BibitemOpen
  \bibfield  {author} {\bibinfo {author} {\bibfnamefont {X.-J.}\ \bibnamefont
  {Liu}}, \bibinfo {author} {\bibfnamefont {K.~T.}\ \bibnamefont {Law}}, \ and\
  \bibinfo {author} {\bibfnamefont {T.~K.}\ \bibnamefont {Ng}},\ }\href
  {\doibase 10.1103/PhysRevLett.112.086401} {\bibfield  {journal} {\bibinfo
  {journal} {Phys. Rev. Lett.}\ }\textbf {\bibinfo {volume} {112}},\ \bibinfo
  {pages} {086401} (\bibinfo {year} {2014})}\BibitemShut {NoStop}%
\bibitem [{\citenamefont {Wang}\ \emph {et~al.}(2010)\citenamefont {Wang},
  \citenamefont {Gao}, \citenamefont {Jian},\ and\ \citenamefont
  {Zhai}}]{wang2010spin}%
  \BibitemOpen
  \bibfield  {author} {\bibinfo {author} {\bibfnamefont {C.}~\bibnamefont
  {Wang}}, \bibinfo {author} {\bibfnamefont {C.}~\bibnamefont {Gao}}, \bibinfo
  {author} {\bibfnamefont {C.-M.}\ \bibnamefont {Jian}}, \ and\ \bibinfo
  {author} {\bibfnamefont {H.}~\bibnamefont {Zhai}},\ }\href {\doibase
  10.1103/PhysRevLett.105.160403} {\bibfield  {journal} {\bibinfo  {journal}
  {Phys. Rev. Lett.}\ }\textbf {\bibinfo {volume} {105}},\ \bibinfo {pages}
  {160403} (\bibinfo {year} {2010})}\BibitemShut {NoStop}%
\bibitem [{\citenamefont {Ho}\ and\ \citenamefont {Zhang}(2011)}]{ho2011bose}%
  \BibitemOpen
  \bibfield  {author} {\bibinfo {author} {\bibfnamefont {T.-L.}\ \bibnamefont
  {Ho}}\ and\ \bibinfo {author} {\bibfnamefont {S.}~\bibnamefont {Zhang}},\
  }\href {\doibase 10.1103/PhysRevLett.107.150403} {\bibfield  {journal}
  {\bibinfo  {journal} {Phys. Rev. Lett.}\ }\textbf {\bibinfo {volume} {107}},\
  \bibinfo {pages} {150403} (\bibinfo {year} {2011})}\BibitemShut {NoStop}%
\bibitem [{\citenamefont {Li}\ \emph {et~al.}(2012{\natexlab{a}})\citenamefont
  {Li}, \citenamefont {Pitaevskii},\ and\ \citenamefont
  {Stringari}}]{li2012quantum}%
  \BibitemOpen
  \bibfield  {author} {\bibinfo {author} {\bibfnamefont {Y.}~\bibnamefont
  {Li}}, \bibinfo {author} {\bibfnamefont {L.~P.}\ \bibnamefont {Pitaevskii}},
  \ and\ \bibinfo {author} {\bibfnamefont {S.}~\bibnamefont {Stringari}},\
  }\href {\doibase 10.1103/PhysRevLett.108.225301} {\bibfield  {journal}
  {\bibinfo  {journal} {Phys. Rev. Lett.}\ }\textbf {\bibinfo {volume} {108}},\
  \bibinfo {pages} {225301} (\bibinfo {year} {2012}{\natexlab{a}})}\BibitemShut
  {NoStop}%
\bibitem [{\citenamefont {Hu}\ \emph {et~al.}(2012)\citenamefont {Hu},
  \citenamefont {Ramachandhran}, \citenamefont {Pu},\ and\ \citenamefont
  {Liu}}]{hu2012spin}%
  \BibitemOpen
  \bibfield  {author} {\bibinfo {author} {\bibfnamefont {H.}~\bibnamefont
  {Hu}}, \bibinfo {author} {\bibfnamefont {B.}~\bibnamefont {Ramachandhran}},
  \bibinfo {author} {\bibfnamefont {H.}~\bibnamefont {Pu}}, \ and\ \bibinfo
  {author} {\bibfnamefont {X.-J.}\ \bibnamefont {Liu}},\ }\href {\doibase
  10.1103/PhysRevLett.108.010402} {\bibfield  {journal} {\bibinfo  {journal}
  {Phys. Rev. Lett.}\ }\textbf {\bibinfo {volume} {108}},\ \bibinfo {pages}
  {010402} (\bibinfo {year} {2012})}\BibitemShut {NoStop}%
\bibitem [{\citenamefont {Bloch}\ \emph {et~al.}(2008)\citenamefont {Bloch},
  \citenamefont {Dalibard},\ and\ \citenamefont {Zwerger}}]{bloch2008many}%
  \BibitemOpen
  \bibfield  {author} {\bibinfo {author} {\bibfnamefont {I.}~\bibnamefont
  {Bloch}}, \bibinfo {author} {\bibfnamefont {J.}~\bibnamefont {Dalibard}}, \
  and\ \bibinfo {author} {\bibfnamefont {W.}~\bibnamefont {Zwerger}},\ }\href
  {\doibase 10.1103/RevModPhys.80.885} {\bibfield  {journal} {\bibinfo
  {journal} {Rev. Mod. Phys.}\ }\textbf {\bibinfo {volume} {80}},\ \bibinfo
  {pages} {885} (\bibinfo {year} {2008})}\BibitemShut {NoStop}%
\bibitem [{\citenamefont {Liu}\ \emph {et~al.}(2009)\citenamefont {Liu},
  \citenamefont {Borunda}, \citenamefont {Liu},\ and\ \citenamefont
  {Sinova}}]{liu2009effect}%
  \BibitemOpen
  \bibfield  {author} {\bibinfo {author} {\bibfnamefont {X.-J.}\ \bibnamefont
  {Liu}}, \bibinfo {author} {\bibfnamefont {M.~F.}\ \bibnamefont {Borunda}},
  \bibinfo {author} {\bibfnamefont {X.}~\bibnamefont {Liu}}, \ and\ \bibinfo
  {author} {\bibfnamefont {J.}~\bibnamefont {Sinova}},\ }\href {\doibase
  10.1103/PhysRevLett.102.046402} {\bibfield  {journal} {\bibinfo  {journal}
  {Phys. Rev. Lett.}\ }\textbf {\bibinfo {volume} {102}},\ \bibinfo {pages}
  {046402} (\bibinfo {year} {2009})}\BibitemShut {NoStop}%
\bibitem [{\citenamefont {Spielman}(2009)}]{spielman2009raman}%
  \BibitemOpen
  \bibfield  {author} {\bibinfo {author} {\bibfnamefont {I.~B.}\ \bibnamefont
  {Spielman}},\ }\href {\doibase 10.1103/PhysRevA.79.063613} {\bibfield
  {journal} {\bibinfo  {journal} {Phys. Rev. A}\ }\textbf {\bibinfo {volume}
  {79}},\ \bibinfo {pages} {063613} (\bibinfo {year} {2009})}\BibitemShut
  {NoStop}%
\bibitem [{\citenamefont {Lin}\ \emph {et~al.}(2011)\citenamefont {Lin},
  \citenamefont {Jimenez-Garcia},\ and\ \citenamefont
  {Spielman}}]{lin2011spin}%
  \BibitemOpen
  \bibfield  {author} {\bibinfo {author} {\bibfnamefont {Y.-J.}\ \bibnamefont
  {Lin}}, \bibinfo {author} {\bibfnamefont {K.}~\bibnamefont {Jimenez-Garcia}},
  \ and\ \bibinfo {author} {\bibfnamefont {I.~B.}\ \bibnamefont {Spielman}},\
  }\href {\doibase 10.1038/nature09887} {\bibfield  {journal} {\bibinfo
  {journal} {Nature}\ }\textbf {\bibinfo {volume} {471}},\ \bibinfo {pages}
  {83} (\bibinfo {year} {2011})}\BibitemShut {NoStop}%
\bibitem [{\citenamefont {Wang}\ \emph {et~al.}(2012)\citenamefont {Wang},
  \citenamefont {Yu}, \citenamefont {Fu}, \citenamefont {Miao}, \citenamefont
  {Huang}, \citenamefont {Chai}, \citenamefont {Zhai},\ and\ \citenamefont
  {Zhang}}]{wang2012spin}%
  \BibitemOpen
  \bibfield  {author} {\bibinfo {author} {\bibfnamefont {P.}~\bibnamefont
  {Wang}}, \bibinfo {author} {\bibfnamefont {Z.-Q.}\ \bibnamefont {Yu}},
  \bibinfo {author} {\bibfnamefont {Z.}~\bibnamefont {Fu}}, \bibinfo {author}
  {\bibfnamefont {J.}~\bibnamefont {Miao}}, \bibinfo {author} {\bibfnamefont
  {L.}~\bibnamefont {Huang}}, \bibinfo {author} {\bibfnamefont
  {S.}~\bibnamefont {Chai}}, \bibinfo {author} {\bibfnamefont {H.}~\bibnamefont
  {Zhai}}, \ and\ \bibinfo {author} {\bibfnamefont {J.}~\bibnamefont {Zhang}},\
  }\href {\doibase 10.1103/PhysRevLett.109.095301} {\bibfield  {journal}
  {\bibinfo  {journal} {Phys. Rev. Lett.}\ }\textbf {\bibinfo {volume} {109}},\
  \bibinfo {pages} {095301} (\bibinfo {year} {2012})}\BibitemShut {NoStop}%
\bibitem [{\citenamefont {Cheuk}\ \emph {et~al.}(2012)\citenamefont {Cheuk},
  \citenamefont {Sommer}, \citenamefont {Hadzibabic}, \citenamefont {Yefsah},
  \citenamefont {Bakr},\ and\ \citenamefont {Zwierlein}}]{cheuk2012spin}%
  \BibitemOpen
  \bibfield  {author} {\bibinfo {author} {\bibfnamefont {L.~W.}\ \bibnamefont
  {Cheuk}}, \bibinfo {author} {\bibfnamefont {A.~T.}\ \bibnamefont {Sommer}},
  \bibinfo {author} {\bibfnamefont {Z.}~\bibnamefont {Hadzibabic}}, \bibinfo
  {author} {\bibfnamefont {T.}~\bibnamefont {Yefsah}}, \bibinfo {author}
  {\bibfnamefont {W.~S.}\ \bibnamefont {Bakr}}, \ and\ \bibinfo {author}
  {\bibfnamefont {M.~W.}\ \bibnamefont {Zwierlein}},\ }\href {\doibase
  10.1103/PhysRevLett.109.095302} {\bibfield  {journal} {\bibinfo  {journal}
  {Phys. Rev. Lett.}\ }\textbf {\bibinfo {volume} {109}},\ \bibinfo {pages}
  {095302} (\bibinfo {year} {2012})}\BibitemShut {NoStop}%
\bibitem [{\citenamefont {Dalibard}\ \emph {et~al.}(2011)\citenamefont
  {Dalibard}, \citenamefont {Gerbier}, \citenamefont
  {Juzeli\ifmmode~\bar{u}\else \={u}\fi{}nas},\ and\ \citenamefont
  {\"Ohberg}}]{dalibard2011colloquium}%
  \BibitemOpen
  \bibfield  {author} {\bibinfo {author} {\bibfnamefont {J.}~\bibnamefont
  {Dalibard}}, \bibinfo {author} {\bibfnamefont {F.}~\bibnamefont {Gerbier}},
  \bibinfo {author} {\bibfnamefont {G.}~\bibnamefont
  {Juzeli\ifmmode~\bar{u}\else \={u}\fi{}nas}}, \ and\ \bibinfo {author}
  {\bibfnamefont {P.}~\bibnamefont {\"Ohberg}},\ }\href {\doibase
  10.1103/RevModPhys.83.1523} {\bibfield  {journal} {\bibinfo  {journal} {Rev.
  Mod. Phys.}\ }\textbf {\bibinfo {volume} {83}},\ \bibinfo {pages} {1523}
  (\bibinfo {year} {2011})}\BibitemShut {NoStop}%
\bibitem [{\citenamefont {Goldman}\ \emph {et~al.}(2014)\citenamefont
  {Goldman}, \citenamefont {Juzeli\={u}nas}, \citenamefont {\"Ohberg},\ and\
  \citenamefont {Spielman}}]{goldman2014light}%
  \BibitemOpen
  \bibfield  {author} {\bibinfo {author} {\bibfnamefont {N.}~\bibnamefont
  {Goldman}}, \bibinfo {author} {\bibfnamefont {G.}~\bibnamefont
  {Juzeli\={u}nas}}, \bibinfo {author} {\bibfnamefont {P.}~\bibnamefont
  {\"Ohberg}}, \ and\ \bibinfo {author} {\bibfnamefont {I.~B.}\ \bibnamefont
  {Spielman}},\ }\href {http://stacks.iop.org/0034-4885/77/i=12/a=126401}
  {\bibfield  {journal} {\bibinfo  {journal} {Reports on Progress in Physics}\
  }\textbf {\bibinfo {volume} {77}},\ \bibinfo {pages} {126401} (\bibinfo
  {year} {2014})}\BibitemShut {NoStop}%
\bibitem [{\citenamefont {Wu}\ \emph {et~al.}(2016)\citenamefont {Wu},
  \citenamefont {Zhang}, \citenamefont {Sun}, \citenamefont {Xu}, \citenamefont
  {Wang}, \citenamefont {Ji}, \citenamefont {Deng}, \citenamefont {Chen},
  \citenamefont {Liu},\ and\ \citenamefont {Pan}}]{wu2016realization}%
  \BibitemOpen
  \bibfield  {author} {\bibinfo {author} {\bibfnamefont {Z.}~\bibnamefont
  {Wu}}, \bibinfo {author} {\bibfnamefont {L.}~\bibnamefont {Zhang}}, \bibinfo
  {author} {\bibfnamefont {W.}~\bibnamefont {Sun}}, \bibinfo {author}
  {\bibfnamefont {X.-T.}\ \bibnamefont {Xu}}, \bibinfo {author} {\bibfnamefont
  {B.-Z.}\ \bibnamefont {Wang}}, \bibinfo {author} {\bibfnamefont {S.-C.}\
  \bibnamefont {Ji}}, \bibinfo {author} {\bibfnamefont {Y.}~\bibnamefont
  {Deng}}, \bibinfo {author} {\bibfnamefont {S.}~\bibnamefont {Chen}}, \bibinfo
  {author} {\bibfnamefont {X.-J.}\ \bibnamefont {Liu}}, \ and\ \bibinfo
  {author} {\bibfnamefont {J.-W.}\ \bibnamefont {Pan}},\ }\href {\doibase
  10.1126/science.aaf6689} {\bibfield  {journal} {\bibinfo  {journal}
  {Science}\ }\textbf {\bibinfo {volume} {354}},\ \bibinfo {pages} {83}
  (\bibinfo {year} {2016})}\BibitemShut {NoStop}%
\bibitem [{\citenamefont {Sun}\ \emph {et~al.}(2017)\citenamefont {Sun},
  \citenamefont {Wang}, \citenamefont {Xu}, \citenamefont {Yi}, \citenamefont
  {Zhang}, \citenamefont {Wu}, \citenamefont {Deng}, \citenamefont {Liu},
  \citenamefont {Chen},\ and\ \citenamefont {Pan}}]{sun2017long}%
  \BibitemOpen
  \bibfield  {author} {\bibinfo {author} {\bibfnamefont {W.}~\bibnamefont
  {Sun}}, \bibinfo {author} {\bibfnamefont {B.-Z.}\ \bibnamefont {Wang}},
  \bibinfo {author} {\bibfnamefont {X.-T.}\ \bibnamefont {Xu}}, \bibinfo
  {author} {\bibfnamefont {C.-R.}\ \bibnamefont {Yi}}, \bibinfo {author}
  {\bibfnamefont {L.}~\bibnamefont {Zhang}}, \bibinfo {author} {\bibfnamefont
  {Z.}~\bibnamefont {Wu}}, \bibinfo {author} {\bibfnamefont {Y.}~\bibnamefont
  {Deng}}, \bibinfo {author} {\bibfnamefont {X.-J.}\ \bibnamefont {Liu}},
  \bibinfo {author} {\bibfnamefont {S.}~\bibnamefont {Chen}}, \ and\ \bibinfo
  {author} {\bibfnamefont {J.-W.}\ \bibnamefont {Pan}},\ }\href
  {https://arxiv.org/abs/1710.00717} {\bibfield  {journal} {\bibinfo  {journal}
  {arXiv preprint arXiv:1710.00717}\ } (\bibinfo {year} {2017})}\BibitemShut
  {NoStop}%
\bibitem [{\citenamefont {Huang}\ \emph {et~al.}(2016)\citenamefont {Huang},
  \citenamefont {Meng}, \citenamefont {Wang}, \citenamefont {Peng},
  \citenamefont {Zhang}, \citenamefont {Chen}, \citenamefont {Li},
  \citenamefont {Zhou},\ and\ \citenamefont {Zhang}}]{huang2016experimental}%
  \BibitemOpen
  \bibfield  {author} {\bibinfo {author} {\bibfnamefont {L.}~\bibnamefont
  {Huang}}, \bibinfo {author} {\bibfnamefont {Z.}~\bibnamefont {Meng}},
  \bibinfo {author} {\bibfnamefont {P.}~\bibnamefont {Wang}}, \bibinfo {author}
  {\bibfnamefont {P.}~\bibnamefont {Peng}}, \bibinfo {author} {\bibfnamefont
  {S.-L.}\ \bibnamefont {Zhang}}, \bibinfo {author} {\bibfnamefont
  {L.}~\bibnamefont {Chen}}, \bibinfo {author} {\bibfnamefont {D.}~\bibnamefont
  {Li}}, \bibinfo {author} {\bibfnamefont {Q.}~\bibnamefont {Zhou}}, \ and\
  \bibinfo {author} {\bibfnamefont {J.}~\bibnamefont {Zhang}},\ }\href
  {http://dx.doi.org/10.1038/nphys3672} {\bibfield  {journal} {\bibinfo
  {journal} {Nat Phys}\ }\textbf {\bibinfo {volume} {12}},\ \bibinfo {pages}
  {540} (\bibinfo {year} {2016})}\BibitemShut {NoStop}%
\bibitem [{\citenamefont {Li}\ \emph {et~al.}(2012{\natexlab{b}})\citenamefont
  {Li}, \citenamefont {Martone},\ and\ \citenamefont {Stringari}}]{li2012sum}%
  \BibitemOpen
  \bibfield  {author} {\bibinfo {author} {\bibfnamefont {Y.}~\bibnamefont
  {Li}}, \bibinfo {author} {\bibfnamefont {G.~I.}\ \bibnamefont {Martone}}, \
  and\ \bibinfo {author} {\bibfnamefont {S.}~\bibnamefont {Stringari}},\ }\href
  {http://stacks.iop.org/0295-5075/99/i=5/a=56008} {\bibfield  {journal}
  {\bibinfo  {journal} {EPL (Europhysics Letters)}\ }\textbf {\bibinfo {volume}
  {99}},\ \bibinfo {pages} {56008} (\bibinfo {year}
  {2012}{\natexlab{b}})}\BibitemShut {NoStop}%
\bibitem [{\citenamefont {Martone}\ \emph {et~al.}(2012)\citenamefont
  {Martone}, \citenamefont {Li}, \citenamefont {Pitaevskii},\ and\
  \citenamefont {Stringari}}]{martone2012anisotropic}%
  \BibitemOpen
  \bibfield  {author} {\bibinfo {author} {\bibfnamefont {G.~I.}\ \bibnamefont
  {Martone}}, \bibinfo {author} {\bibfnamefont {Y.}~\bibnamefont {Li}},
  \bibinfo {author} {\bibfnamefont {L.~P.}\ \bibnamefont {Pitaevskii}}, \ and\
  \bibinfo {author} {\bibfnamefont {S.}~\bibnamefont {Stringari}},\ }\href
  {\doibase 10.1103/PhysRevA.86.063621} {\bibfield  {journal} {\bibinfo
  {journal} {Phys. Rev. A}\ }\textbf {\bibinfo {volume} {86}},\ \bibinfo
  {pages} {063621} (\bibinfo {year} {2012})}\BibitemShut {NoStop}%
\bibitem [{\citenamefont {Zheng}\ \emph {et~al.}(2013)\citenamefont {Zheng},
  \citenamefont {Yu}, \citenamefont {Cui},\ and\ \citenamefont
  {Zhai}}]{zheng2013properties}%
  \BibitemOpen
  \bibfield  {author} {\bibinfo {author} {\bibfnamefont {W.}~\bibnamefont
  {Zheng}}, \bibinfo {author} {\bibfnamefont {Z.-Q.}\ \bibnamefont {Yu}},
  \bibinfo {author} {\bibfnamefont {X.}~\bibnamefont {Cui}}, \ and\ \bibinfo
  {author} {\bibfnamefont {H.}~\bibnamefont {Zhai}},\ }\href {\doibase
  http://iopscience.iop.org/article/10.1088/0953-4075/46/13/134007/meta}
  {\bibfield  {journal} {\bibinfo  {journal} {Journal of Physics B: Atomic,
  Molecular and Optical Physics}\ }\textbf {\bibinfo {volume} {46}},\ \bibinfo
  {pages} {134007} (\bibinfo {year} {2013})}\BibitemShut {NoStop}%
\bibitem [{\citenamefont {Zhang}\ \emph {et~al.}(2012)\citenamefont {Zhang},
  \citenamefont {Ji}, \citenamefont {Chen}, \citenamefont {Zhang},
  \citenamefont {Du}, \citenamefont {Yan}, \citenamefont {Pan}, \citenamefont
  {Zhao}, \citenamefont {Deng}, \citenamefont {Zhai}, \citenamefont {Chen},\
  and\ \citenamefont {Pan}}]{zhang2012collective}%
  \BibitemOpen
  \bibfield  {author} {\bibinfo {author} {\bibfnamefont {J.-Y.}\ \bibnamefont
  {Zhang}}, \bibinfo {author} {\bibfnamefont {S.-C.}\ \bibnamefont {Ji}},
  \bibinfo {author} {\bibfnamefont {Z.}~\bibnamefont {Chen}}, \bibinfo {author}
  {\bibfnamefont {L.}~\bibnamefont {Zhang}}, \bibinfo {author} {\bibfnamefont
  {Z.-D.}\ \bibnamefont {Du}}, \bibinfo {author} {\bibfnamefont
  {B.}~\bibnamefont {Yan}}, \bibinfo {author} {\bibfnamefont {G.-S.}\
  \bibnamefont {Pan}}, \bibinfo {author} {\bibfnamefont {B.}~\bibnamefont
  {Zhao}}, \bibinfo {author} {\bibfnamefont {Y.-J.}\ \bibnamefont {Deng}},
  \bibinfo {author} {\bibfnamefont {H.}~\bibnamefont {Zhai}}, \bibinfo {author}
  {\bibfnamefont {S.}~\bibnamefont {Chen}}, \ and\ \bibinfo {author}
  {\bibfnamefont {J.-W.}\ \bibnamefont {Pan}},\ }\href {\doibase
  10.1103/PhysRevLett.109.115301} {\bibfield  {journal} {\bibinfo  {journal}
  {Phys. Rev. Lett.}\ }\textbf {\bibinfo {volume} {109}},\ \bibinfo {pages}
  {115301} (\bibinfo {year} {2012})}\BibitemShut {NoStop}%
\bibitem [{\citenamefont {Ji}\ \emph {et~al.}(2014)\citenamefont {Ji},
  \citenamefont {Zhang}, \citenamefont {Zhang}, \citenamefont {Du},
  \citenamefont {Zheng}, \citenamefont {Deng}, \citenamefont {Zhai},
  \citenamefont {Chen},\ and\ \citenamefont {Pan}}]{ji2014experimental}%
  \BibitemOpen
  \bibfield  {author} {\bibinfo {author} {\bibfnamefont {S.-C.}\ \bibnamefont
  {Ji}}, \bibinfo {author} {\bibfnamefont {J.-Y.}\ \bibnamefont {Zhang}},
  \bibinfo {author} {\bibfnamefont {L.}~\bibnamefont {Zhang}}, \bibinfo
  {author} {\bibfnamefont {Z.-D.}\ \bibnamefont {Du}}, \bibinfo {author}
  {\bibfnamefont {W.}~\bibnamefont {Zheng}}, \bibinfo {author} {\bibfnamefont
  {Y.-J.}\ \bibnamefont {Deng}}, \bibinfo {author} {\bibfnamefont
  {H.}~\bibnamefont {Zhai}}, \bibinfo {author} {\bibfnamefont {S.}~\bibnamefont
  {Chen}}, \ and\ \bibinfo {author} {\bibfnamefont {J.-W.}\ \bibnamefont
  {Pan}},\ }\href {http://dx.doi.org/10.1038/nphys2905} {\bibfield  {journal}
  {\bibinfo  {journal} {Nat Phys}\ }\textbf {\bibinfo {volume} {10}},\ \bibinfo
  {pages} {314} (\bibinfo {year} {2014})}\BibitemShut {NoStop}%
\bibitem [{\citenamefont {Ji}\ \emph {et~al.}(2015)\citenamefont {Ji},
  \citenamefont {Zhang}, \citenamefont {Xu}, \citenamefont {Wu}, \citenamefont
  {Deng}, \citenamefont {Chen},\ and\ \citenamefont {Pan}}]{ji2015softening}%
  \BibitemOpen
  \bibfield  {author} {\bibinfo {author} {\bibfnamefont {S.-C.}\ \bibnamefont
  {Ji}}, \bibinfo {author} {\bibfnamefont {L.}~\bibnamefont {Zhang}}, \bibinfo
  {author} {\bibfnamefont {X.-T.}\ \bibnamefont {Xu}}, \bibinfo {author}
  {\bibfnamefont {Z.}~\bibnamefont {Wu}}, \bibinfo {author} {\bibfnamefont
  {Y.}~\bibnamefont {Deng}}, \bibinfo {author} {\bibfnamefont {S.}~\bibnamefont
  {Chen}}, \ and\ \bibinfo {author} {\bibfnamefont {J.-W.}\ \bibnamefont
  {Pan}},\ }\href {\doibase 10.1103/PhysRevLett.114.105301} {\bibfield
  {journal} {\bibinfo  {journal} {Phys. Rev. Lett.}\ }\textbf {\bibinfo
  {volume} {114}},\ \bibinfo {pages} {105301} (\bibinfo {year}
  {2015})}\BibitemShut {NoStop}%
\bibitem [{\citenamefont {Ozawa}\ and\ \citenamefont
  {Baym}(2012)}]{ozawa2012stability}%
  \BibitemOpen
  \bibfield  {author} {\bibinfo {author} {\bibfnamefont {T.}~\bibnamefont
  {Ozawa}}\ and\ \bibinfo {author} {\bibfnamefont {G.}~\bibnamefont {Baym}},\
  }\href {\doibase 10.1103/PhysRevLett.109.025301} {\bibfield  {journal}
  {\bibinfo  {journal} {Phys. Rev. Lett.}\ }\textbf {\bibinfo {volume} {109}},\
  \bibinfo {pages} {025301} (\bibinfo {year} {2012})}\BibitemShut {NoStop}%
\bibitem [{\citenamefont {Cui}\ and\ \citenamefont
  {Zhou}(2013)}]{cui2013enhancement}%
  \BibitemOpen
  \bibfield  {author} {\bibinfo {author} {\bibfnamefont {X.}~\bibnamefont
  {Cui}}\ and\ \bibinfo {author} {\bibfnamefont {Q.}~\bibnamefont {Zhou}},\
  }\href {\doibase 10.1103/PhysRevA.87.031604} {\bibfield  {journal} {\bibinfo
  {journal} {Phys. Rev. A}\ }\textbf {\bibinfo {volume} {87}},\ \bibinfo
  {pages} {031604} (\bibinfo {year} {2013})}\BibitemShut {NoStop}%
\bibitem [{\citenamefont {Liao}\ \emph {et~al.}(2014)\citenamefont {Liao},
  \citenamefont {Huang}, \citenamefont {Lin},\ and\ \citenamefont
  {Fialko}}]{liao2014spin}%
  \BibitemOpen
  \bibfield  {author} {\bibinfo {author} {\bibfnamefont {R.}~\bibnamefont
  {Liao}}, \bibinfo {author} {\bibfnamefont {Z.-G.}\ \bibnamefont {Huang}},
  \bibinfo {author} {\bibfnamefont {X.-M.}\ \bibnamefont {Lin}}, \ and\
  \bibinfo {author} {\bibfnamefont {O.}~\bibnamefont {Fialko}},\ }\href
  {\doibase 10.1103/PhysRevA.89.063614} {\bibfield  {journal} {\bibinfo
  {journal} {Phys. Rev. A}\ }\textbf {\bibinfo {volume} {89}},\ \bibinfo
  {pages} {063614} (\bibinfo {year} {2014})}\BibitemShut {NoStop}%
\bibitem [{\citenamefont {Chen}\ \emph
  {et~al.}(2017{\natexlab{a}})\citenamefont {Chen}, \citenamefont {Liu},\ and\
  \citenamefont {Hu}}]{chen2017quantum}%
  \BibitemOpen
  \bibfield  {author} {\bibinfo {author} {\bibfnamefont {X.-L.}\ \bibnamefont
  {Chen}}, \bibinfo {author} {\bibfnamefont {X.-J.}\ \bibnamefont {Liu}}, \
  and\ \bibinfo {author} {\bibfnamefont {H.}~\bibnamefont {Hu}},\ }\href
  {\doibase 10.1103/PhysRevA.96.013625} {\bibfield  {journal} {\bibinfo
  {journal} {Phys. Rev. A}\ }\textbf {\bibinfo {volume} {96}},\ \bibinfo
  {pages} {013625} (\bibinfo {year} {2017}{\natexlab{a}})}\BibitemShut
  {NoStop}%
\bibitem [{\citenamefont {Khamehchi}\ \emph {et~al.}(2014)\citenamefont
  {Khamehchi}, \citenamefont {Zhang}, \citenamefont {Hamner}, \citenamefont
  {Busch},\ and\ \citenamefont {Engels}}]{khamehchi2014measurement}%
  \BibitemOpen
  \bibfield  {author} {\bibinfo {author} {\bibfnamefont {M.~A.}\ \bibnamefont
  {Khamehchi}}, \bibinfo {author} {\bibfnamefont {Y.}~\bibnamefont {Zhang}},
  \bibinfo {author} {\bibfnamefont {C.}~\bibnamefont {Hamner}}, \bibinfo
  {author} {\bibfnamefont {T.}~\bibnamefont {Busch}}, \ and\ \bibinfo {author}
  {\bibfnamefont {P.}~\bibnamefont {Engels}},\ }\href {\doibase
  10.1103/PhysRevA.90.063624} {\bibfield  {journal} {\bibinfo  {journal} {Phys.
  Rev. A}\ }\textbf {\bibinfo {volume} {90}},\ \bibinfo {pages} {063624}
  (\bibinfo {year} {2014})}\BibitemShut {NoStop}%
\bibitem [{\citenamefont {Chen}\ \emph
  {et~al.}(2017{\natexlab{b}})\citenamefont {Chen}, \citenamefont {Pu},
  \citenamefont {Yu},\ and\ \citenamefont {Zhang}}]{chen2017collective}%
  \BibitemOpen
  \bibfield  {author} {\bibinfo {author} {\bibfnamefont {L.}~\bibnamefont
  {Chen}}, \bibinfo {author} {\bibfnamefont {H.}~\bibnamefont {Pu}}, \bibinfo
  {author} {\bibfnamefont {Z.-Q.}\ \bibnamefont {Yu}}, \ and\ \bibinfo {author}
  {\bibfnamefont {Y.}~\bibnamefont {Zhang}},\ }\href {\doibase
  10.1103/PhysRevA.95.033616} {\bibfield  {journal} {\bibinfo  {journal} {Phys.
  Rev. A}\ }\textbf {\bibinfo {volume} {95}},\ \bibinfo {pages} {033616}
  (\bibinfo {year} {2017}{\natexlab{b}})}\BibitemShut {NoStop}%
\bibitem [{\citenamefont {Zhu}\ \emph {et~al.}(2012)\citenamefont {Zhu},
  \citenamefont {Zhang},\ and\ \citenamefont {Wu}}]{zhu2012exotic}%
  \BibitemOpen
  \bibfield  {author} {\bibinfo {author} {\bibfnamefont {Q.}~\bibnamefont
  {Zhu}}, \bibinfo {author} {\bibfnamefont {C.}~\bibnamefont {Zhang}}, \ and\
  \bibinfo {author} {\bibfnamefont {B.}~\bibnamefont {Wu}},\ }\href
  {http://stacks.iop.org/0295-5075/100/i=5/a=50003} {\bibfield  {journal}
  {\bibinfo  {journal} {EPL (Europhysics Letters)}\ }\textbf {\bibinfo {volume}
  {100}},\ \bibinfo {pages} {50003} (\bibinfo {year} {2012})}\BibitemShut
  {NoStop}%
\bibitem [{\citenamefont {Zhou}\ and\ \citenamefont
  {Zhang}(2012)}]{zhou2012opposite}%
  \BibitemOpen
  \bibfield  {author} {\bibinfo {author} {\bibfnamefont {K.}~\bibnamefont
  {Zhou}}\ and\ \bibinfo {author} {\bibfnamefont {Z.}~\bibnamefont {Zhang}},\
  }\href {\doibase 10.1103/PhysRevLett.108.025301} {\bibfield  {journal}
  {\bibinfo  {journal} {Phys. Rev. Lett.}\ }\textbf {\bibinfo {volume} {108}},\
  \bibinfo {pages} {025301} (\bibinfo {year} {2012})}\BibitemShut {NoStop}%
\bibitem [{\citenamefont {Yu}(2017)}]{yu2017landau}%
  \BibitemOpen
  \bibfield  {author} {\bibinfo {author} {\bibfnamefont {Z.-Q.}\ \bibnamefont
  {Yu}},\ }\href {\doibase 10.1103/PhysRevA.95.033618} {\bibfield  {journal}
  {\bibinfo  {journal} {Phys. Rev. A}\ }\textbf {\bibinfo {volume} {95}},\
  \bibinfo {pages} {033618} (\bibinfo {year} {2017})}\BibitemShut {NoStop}%
\bibitem [{\citenamefont {L{\'e}onard}\ \emph {et~al.}(2017)\citenamefont
  {L{\'e}onard}, \citenamefont {Morales}, \citenamefont {Zupancic},
  \citenamefont {Esslinger},\ and\ \citenamefont
  {Donner}}]{leonard2017supersolid}%
  \BibitemOpen
  \bibfield  {author} {\bibinfo {author} {\bibfnamefont {J.}~\bibnamefont
  {L{\'e}onard}}, \bibinfo {author} {\bibfnamefont {A.}~\bibnamefont
  {Morales}}, \bibinfo {author} {\bibfnamefont {P.}~\bibnamefont {Zupancic}},
  \bibinfo {author} {\bibfnamefont {T.}~\bibnamefont {Esslinger}}, \ and\
  \bibinfo {author} {\bibfnamefont {T.}~\bibnamefont {Donner}},\ }\href
  {http://dx.doi.org/10.1038/nature21067} {\bibfield  {journal} {\bibinfo
  {journal} {Nature}\ }\textbf {\bibinfo {volume} {543}},\ \bibinfo {pages}
  {87} (\bibinfo {year} {2017})}\BibitemShut {NoStop}%
\bibitem [{\citenamefont {Li}\ \emph {et~al.}(2017)\citenamefont {Li},
  \citenamefont {Lee}, \citenamefont {Huang}, \citenamefont {Burchesky},
  \citenamefont {Shteynas}, \citenamefont {Top}, \citenamefont {Jamison},\ and\
  \citenamefont {Ketterle}}]{li2017stripe}%
  \BibitemOpen
  \bibfield  {author} {\bibinfo {author} {\bibfnamefont {J.-R.}\ \bibnamefont
  {Li}}, \bibinfo {author} {\bibfnamefont {J.}~\bibnamefont {Lee}}, \bibinfo
  {author} {\bibfnamefont {W.}~\bibnamefont {Huang}}, \bibinfo {author}
  {\bibfnamefont {S.}~\bibnamefont {Burchesky}}, \bibinfo {author}
  {\bibfnamefont {B.}~\bibnamefont {Shteynas}}, \bibinfo {author}
  {\bibfnamefont {F.~{\c{C}}.}\ \bibnamefont {Top}}, \bibinfo {author}
  {\bibfnamefont {A.~O.}\ \bibnamefont {Jamison}}, \ and\ \bibinfo {author}
  {\bibfnamefont {W.}~\bibnamefont {Ketterle}},\ }\href
  {http://dx.doi.org/10.1038/nature21431} {\bibfield  {journal} {\bibinfo
  {journal} {Nature}\ }\textbf {\bibinfo {volume} {543}},\ \bibinfo {pages}
  {91} (\bibinfo {year} {2017})}\BibitemShut {NoStop}%
\bibitem [{\citenamefont {Li}\ \emph {et~al.}(2013)\citenamefont {Li},
  \citenamefont {Martone}, \citenamefont {Pitaevskii},\ and\ \citenamefont
  {Stringari}}]{li2013superstripes}%
  \BibitemOpen
  \bibfield  {author} {\bibinfo {author} {\bibfnamefont {Y.}~\bibnamefont
  {Li}}, \bibinfo {author} {\bibfnamefont {G.~I.}\ \bibnamefont {Martone}},
  \bibinfo {author} {\bibfnamefont {L.~P.}\ \bibnamefont {Pitaevskii}}, \ and\
  \bibinfo {author} {\bibfnamefont {S.}~\bibnamefont {Stringari}},\ }\href
  {\doibase 10.1103/PhysRevLett.110.235302} {\bibfield  {journal} {\bibinfo
  {journal} {Phys. Rev. Lett.}\ }\textbf {\bibinfo {volume} {110}},\ \bibinfo
  {pages} {235302} (\bibinfo {year} {2013})}\BibitemShut {NoStop}%
\bibitem [{\citenamefont {Martone}\ \emph {et~al.}(2014)\citenamefont
  {Martone}, \citenamefont {Li},\ and\ \citenamefont
  {Stringari}}]{martone2014approach}%
  \BibitemOpen
  \bibfield  {author} {\bibinfo {author} {\bibfnamefont {G.~I.}\ \bibnamefont
  {Martone}}, \bibinfo {author} {\bibfnamefont {Y.}~\bibnamefont {Li}}, \ and\
  \bibinfo {author} {\bibfnamefont {S.}~\bibnamefont {Stringari}},\ }\href
  {\doibase 10.1103/PhysRevA.90.041604} {\bibfield  {journal} {\bibinfo
  {journal} {Phys. Rev. A}\ }\textbf {\bibinfo {volume} {90}},\ \bibinfo
  {pages} {041604} (\bibinfo {year} {2014})}\BibitemShut {NoStop}%
\bibitem [{\citenamefont {Li}\ \emph {et~al.}(2016)\citenamefont {Li},
  \citenamefont {Huang}, \citenamefont {Shteynas}, \citenamefont {Burchesky},
  \citenamefont {Top}, \citenamefont {Su}, \citenamefont {Lee}, \citenamefont
  {Jamison},\ and\ \citenamefont {Ketterle}}]{li2016spin}%
  \BibitemOpen
  \bibfield  {author} {\bibinfo {author} {\bibfnamefont {J.}~\bibnamefont
  {Li}}, \bibinfo {author} {\bibfnamefont {W.}~\bibnamefont {Huang}}, \bibinfo
  {author} {\bibfnamefont {B.}~\bibnamefont {Shteynas}}, \bibinfo {author}
  {\bibfnamefont {S.}~\bibnamefont {Burchesky}}, \bibinfo {author}
  {\bibfnamefont {F.~{\c{C}}.}\ \bibnamefont {Top}}, \bibinfo {author}
  {\bibfnamefont {E.}~\bibnamefont {Su}}, \bibinfo {author} {\bibfnamefont
  {J.}~\bibnamefont {Lee}}, \bibinfo {author} {\bibfnamefont {A.~O.}\
  \bibnamefont {Jamison}}, \ and\ \bibinfo {author} {\bibfnamefont
  {W.}~\bibnamefont {Ketterle}},\ }\href {\doibase
  10.1103/PhysRevLett.117.185301} {\bibfield  {journal} {\bibinfo  {journal}
  {Phys. Rev. Lett.}\ }\textbf {\bibinfo {volume} {117}},\ \bibinfo {pages}
  {185301} (\bibinfo {year} {2016})}\BibitemShut {NoStop}%
\bibitem [{\citenamefont {Griffin}(1996)}]{griffin1996conserving}%
  \BibitemOpen
  \bibfield  {author} {\bibinfo {author} {\bibfnamefont {A.}~\bibnamefont
  {Griffin}},\ }\href {\doibase 10.1103/PhysRevB.53.9341} {\bibfield  {journal}
  {\bibinfo  {journal} {Phys. Rev. B}\ }\textbf {\bibinfo {volume} {53}},\
  \bibinfo {pages} {9341} (\bibinfo {year} {1996})}\BibitemShut {NoStop}%
\bibitem [{\citenamefont {Dodd}\ \emph {et~al.}(1998)\citenamefont {Dodd},
  \citenamefont {Edwards}, \citenamefont {Clark},\ and\ \citenamefont
  {Burnett}}]{dodd1998collective}%
  \BibitemOpen
  \bibfield  {author} {\bibinfo {author} {\bibfnamefont {R.~J.}\ \bibnamefont
  {Dodd}}, \bibinfo {author} {\bibfnamefont {M.}~\bibnamefont {Edwards}},
  \bibinfo {author} {\bibfnamefont {C.~W.}\ \bibnamefont {Clark}}, \ and\
  \bibinfo {author} {\bibfnamefont {K.}~\bibnamefont {Burnett}},\ }\href
  {\doibase 10.1103/PhysRevA.57.R32} {\bibfield  {journal} {\bibinfo  {journal}
  {Phys. Rev. A}\ }\textbf {\bibinfo {volume} {57}},\ \bibinfo {pages} {R32}
  (\bibinfo {year} {1998})}\BibitemShut {NoStop}%
\bibitem [{\citenamefont {Buljan}\ \emph {et~al.}(2005)\citenamefont {Buljan},
  \citenamefont {Segev},\ and\ \citenamefont {Vardi}}]{buljan2005incoherent}%
  \BibitemOpen
  \bibfield  {author} {\bibinfo {author} {\bibfnamefont {H.}~\bibnamefont
  {Buljan}}, \bibinfo {author} {\bibfnamefont {M.}~\bibnamefont {Segev}}, \
  and\ \bibinfo {author} {\bibfnamefont {A.}~\bibnamefont {Vardi}},\ }\href
  {\doibase 10.1103/PhysRevLett.95.180401} {\bibfield  {journal} {\bibinfo
  {journal} {Phys. Rev. Lett.}\ }\textbf {\bibinfo {volume} {95}},\ \bibinfo
  {pages} {180401} (\bibinfo {year} {2005})}\BibitemShut {NoStop}%
\bibitem [{\citenamefont {Dalfovo}\ \emph {et~al.}(1999)\citenamefont
  {Dalfovo}, \citenamefont {Giorgini}, \citenamefont {Pitaevskii},\ and\
  \citenamefont {Stringari}}]{dalfovo1999theory}%
  \BibitemOpen
  \bibfield  {author} {\bibinfo {author} {\bibfnamefont {F.}~\bibnamefont
  {Dalfovo}}, \bibinfo {author} {\bibfnamefont {S.}~\bibnamefont {Giorgini}},
  \bibinfo {author} {\bibfnamefont {L.~P.}\ \bibnamefont {Pitaevskii}}, \ and\
  \bibinfo {author} {\bibfnamefont {S.}~\bibnamefont {Stringari}},\ }\href
  {\doibase 10.1103/RevModPhys.71.463} {\bibfield  {journal} {\bibinfo
  {journal} {Rev. Mod. Phys.}\ }\textbf {\bibinfo {volume} {71}},\ \bibinfo
  {pages} {463} (\bibinfo {year} {1999})}\BibitemShut {NoStop}%
\bibitem [{\citenamefont {Pitaevskii}\ and\ \citenamefont
  {Stringari}(2003)}]{Pitaevskii2003Book}%
  \BibitemOpen
  \bibfield  {author} {\bibinfo {author} {\bibfnamefont {L.~P.}\ \bibnamefont
  {Pitaevskii}}\ and\ \bibinfo {author} {\bibfnamefont {S.}~\bibnamefont
  {Stringari}},\ }\href@noop {} {\emph {\bibinfo {title} {Bose-Einstein
  Condensation}}}\ (\bibinfo  {publisher} {Oxford University Press},\ \bibinfo
  {year} {2003})\ Chap.~\bibinfo {chapter} {4}\BibitemShut {NoStop}%
\bibitem [{\citenamefont {Chen}\ \emph {et~al.}(2015)\citenamefont {Chen},
  \citenamefont {Li},\ and\ \citenamefont {Hu}}]{chen2015collective}%
  \BibitemOpen
  \bibfield  {author} {\bibinfo {author} {\bibfnamefont {X.-L.}\ \bibnamefont
  {Chen}}, \bibinfo {author} {\bibfnamefont {Y.}~\bibnamefont {Li}}, \ and\
  \bibinfo {author} {\bibfnamefont {H.}~\bibnamefont {Hu}},\ }\href {\doibase
  10.1103/PhysRevA.91.063631} {\bibfield  {journal} {\bibinfo  {journal} {Phys.
  Rev. A}\ }\textbf {\bibinfo {volume} {91}},\ \bibinfo {pages} {063631}
  (\bibinfo {year} {2015})}\BibitemShut {NoStop}%
\bibitem [{\citenamefont {Zhang}\ \emph {et~al.}(2016)\citenamefont {Zhang},
  \citenamefont {Yu}, \citenamefont {Ng}, \citenamefont {Zhang}, \citenamefont
  {Pitaevskii},\ and\ \citenamefont {Stringari}}]{zhang2016superfluid}%
  \BibitemOpen
  \bibfield  {author} {\bibinfo {author} {\bibfnamefont {Y.-C.}\ \bibnamefont
  {Zhang}}, \bibinfo {author} {\bibfnamefont {Z.-Q.}\ \bibnamefont {Yu}},
  \bibinfo {author} {\bibfnamefont {T.~K.}\ \bibnamefont {Ng}}, \bibinfo
  {author} {\bibfnamefont {S.}~\bibnamefont {Zhang}}, \bibinfo {author}
  {\bibfnamefont {L.}~\bibnamefont {Pitaevskii}}, \ and\ \bibinfo {author}
  {\bibfnamefont {S.}~\bibnamefont {Stringari}},\ }\href {\doibase
  10.1103/PhysRevA.94.033635} {\bibfield  {journal} {\bibinfo  {journal} {Phys.
  Rev. A}\ }\textbf {\bibinfo {volume} {94}},\ \bibinfo {pages} {033635}
  (\bibinfo {year} {2016})}\BibitemShut {NoStop}%
\bibitem [{\citenamefont {Yu}(2014)}]{yu2014equation}%
  \BibitemOpen
  \bibfield  {author} {\bibinfo {author} {\bibfnamefont {Z.-Q.}\ \bibnamefont
  {Yu}},\ }\href {\doibase 10.1103/PhysRevA.90.053608} {\bibfield  {journal}
  {\bibinfo  {journal} {Phys. Rev. A}\ }\textbf {\bibinfo {volume} {90}},\
  \bibinfo {pages} {053608} (\bibinfo {year} {2014})}\BibitemShut {NoStop}%
\bibitem [{Note1()}]{Note1}%
  \BibitemOpen
  \bibinfo {note} {\label {note1}The condition is necessary for the existence
  of the exotic stripe phase, where the more strict one is $E_{\protect \mathrm
  {r}}>2G_{2}+2G_{2}^{2}/G_{1}$ in Ref.~\cite {li2012quantum}.}\BibitemShut
  {Stop}%
\bibitem [{\citenamefont {Fisher}\ \emph {et~al.}(1973)\citenamefont {Fisher},
  \citenamefont {Barber},\ and\ \citenamefont {Jasnow}}]{fisher1973helicity}%
  \BibitemOpen
  \bibfield  {author} {\bibinfo {author} {\bibfnamefont {M.~E.}\ \bibnamefont
  {Fisher}}, \bibinfo {author} {\bibfnamefont {M.~N.}\ \bibnamefont {Barber}},
  \ and\ \bibinfo {author} {\bibfnamefont {D.}~\bibnamefont {Jasnow}},\ }\href
  {\doibase 10.1103/PhysRevA.8.1111} {\bibfield  {journal} {\bibinfo  {journal}
  {Phys. Rev. A}\ }\textbf {\bibinfo {volume} {8}},\ \bibinfo {pages} {1111}
  (\bibinfo {year} {1973})}\BibitemShut {NoStop}%
\bibitem [{\citenamefont {Taylor}\ \emph {et~al.}(2006)\citenamefont {Taylor},
  \citenamefont {Griffin}, \citenamefont {Fukushima},\ and\ \citenamefont
  {Ohashi}}]{taylor2006pairing}%
  \BibitemOpen
  \bibfield  {author} {\bibinfo {author} {\bibfnamefont {E.}~\bibnamefont
  {Taylor}}, \bibinfo {author} {\bibfnamefont {A.}~\bibnamefont {Griffin}},
  \bibinfo {author} {\bibfnamefont {N.}~\bibnamefont {Fukushima}}, \ and\
  \bibinfo {author} {\bibfnamefont {Y.}~\bibnamefont {Ohashi}},\ }\href
  {\doibase 10.1103/PhysRevA.74.063626} {\bibfield  {journal} {\bibinfo
  {journal} {Phys. Rev. A}\ }\textbf {\bibinfo {volume} {74}},\ \bibinfo
  {pages} {063626} (\bibinfo {year} {2006})}\BibitemShut {NoStop}%
\bibitem [{\citenamefont {He}\ \emph {et~al.}(2018)\citenamefont {He},
  \citenamefont {Hu},\ and\ \citenamefont {Liu}}]{he2018realizing}%
  \BibitemOpen
  \bibfield  {author} {\bibinfo {author} {\bibfnamefont {L.}~\bibnamefont
  {He}}, \bibinfo {author} {\bibfnamefont {H.}~\bibnamefont {Hu}}, \ and\
  \bibinfo {author} {\bibfnamefont {X.-J.}\ \bibnamefont {Liu}},\ }\href
  {\doibase 10.1103/PhysRevLett.120.045302} {\bibfield  {journal} {\bibinfo
  {journal} {Phys. Rev. Lett.}\ }\textbf {\bibinfo {volume} {120}},\ \bibinfo
  {pages} {045302} (\bibinfo {year} {2018})}\BibitemShut {NoStop}%
\bibitem [{Note2()}]{Note2}%
  \BibitemOpen
  \bibinfo {note} {In this work, the phase twist is along the direction of SOC,
  i.e., the $x$ direction, or in the perpendicular $y-z$ plane}\BibitemShut
  {NoStop}%
\bibitem [{\citenamefont {Mivehvar}\ and\ \citenamefont
  {Feder}(2015)}]{mivehvar2015enhanced}%
  \BibitemOpen
  \bibfield  {author} {\bibinfo {author} {\bibfnamefont {F.}~\bibnamefont
  {Mivehvar}}\ and\ \bibinfo {author} {\bibfnamefont {D.~L.}\ \bibnamefont
  {Feder}},\ }\href {\doibase 10.1103/PhysRevA.92.023611} {\bibfield  {journal}
  {\bibinfo  {journal} {Phys. Rev. A}\ }\textbf {\bibinfo {volume} {92}},\
  \bibinfo {pages} {023611} (\bibinfo {year} {2015})}\BibitemShut {NoStop}%
\bibitem [{\citenamefont {Abad}\ and\ \citenamefont
  {Recati}(2013)}]{abad2013study}%
  \BibitemOpen
  \bibfield  {author} {\bibinfo {author} {\bibfnamefont {M.}~\bibnamefont
  {Abad}}\ and\ \bibinfo {author} {\bibfnamefont {A.}~\bibnamefont {Recati}},\
  }\href {\doibase 10.1140/epjd/e2013-40053-2} {\bibfield  {journal} {\bibinfo
  {journal} {The European Physical Journal D}\ }\textbf {\bibinfo {volume}
  {67}},\ \bibinfo {pages} {148} (\bibinfo {year} {2013})}\BibitemShut
  {NoStop}%
\bibitem [{\citenamefont {Lopes}\ \emph {et~al.}(2017)\citenamefont {Lopes},
  \citenamefont {Eigen}, \citenamefont {Navon}, \citenamefont {Cl\'ement},
  \citenamefont {Smith},\ and\ \citenamefont {Hadzibabic}}]{lopes2017quantum}%
  \BibitemOpen
  \bibfield  {author} {\bibinfo {author} {\bibfnamefont {R.}~\bibnamefont
  {Lopes}}, \bibinfo {author} {\bibfnamefont {C.}~\bibnamefont {Eigen}},
  \bibinfo {author} {\bibfnamefont {N.}~\bibnamefont {Navon}}, \bibinfo
  {author} {\bibfnamefont {D.}~\bibnamefont {Cl\'ement}}, \bibinfo {author}
  {\bibfnamefont {R.~P.}\ \bibnamefont {Smith}}, \ and\ \bibinfo {author}
  {\bibfnamefont {Z.}~\bibnamefont {Hadzibabic}},\ }\href {\doibase
  10.1103/PhysRevLett.119.190404} {\bibfield  {journal} {\bibinfo  {journal}
  {Phys. Rev. Lett.}\ }\textbf {\bibinfo {volume} {119}},\ \bibinfo {pages}
  {190404} (\bibinfo {year} {2017})}\BibitemShut {NoStop}%
\bibitem [{\citenamefont {Pethick}\ and\ \citenamefont
  {Smith}(2002)}]{pethick2002bose}%
  \BibitemOpen
  \bibfield  {author} {\bibinfo {author} {\bibfnamefont {C.~J.}\ \bibnamefont
  {Pethick}}\ and\ \bibinfo {author} {\bibfnamefont {H.}~\bibnamefont
  {Smith}},\ }\href@noop {} {\emph {\bibinfo {title} {Bose-Einstein
  condensation in dilute gases}}}\ (\bibinfo  {publisher} {Cambridge university
  press},\ \bibinfo {year} {2002})\BibitemShut {NoStop}%
\bibitem [{\citenamefont {Stringari}(2017)}]{stringari2017diffused}%
  \BibitemOpen
  \bibfield  {author} {\bibinfo {author} {\bibfnamefont {S.}~\bibnamefont
  {Stringari}},\ }\href {\doibase 10.1103/PhysRevLett.118.145302} {\bibfield
  {journal} {\bibinfo  {journal} {Phys. Rev. Lett.}\ }\textbf {\bibinfo
  {volume} {118}},\ \bibinfo {pages} {145302} (\bibinfo {year}
  {2017})}\BibitemShut {NoStop}%
\end{thebibliography}%

\end{document}